\documentclass[a4paper,11pt]{article}

\usepackage{amsfonts}
\usepackage{amsmath}
\usepackage{amssymb}
\usepackage{amssymb}
\usepackage[titletoc,toc,title]{appendix}
\usepackage{authblk}
\usepackage{braket}
\usepackage{cancel}
\usepackage[font={footnotesize,it}]{caption}
\usepackage{cite}
\usepackage{comment}
\usepackage{enumitem}
\usepackage{epsfig}
\usepackage{float}
\usepackage[margin=2.5cm]{geometry}
\usepackage{graphicx}
\usepackage{epstopdf}
\usepackage{hyperref}
\usepackage{cleveref}
\usepackage{physics}
\usepackage{ragged2e}
\usepackage{soul}
\usepackage{subcaption}
\usepackage{xcolor}
\usepackage{array}
\usepackage{enumitem}
\usepackage{amssymb}
\newlist{subquestion}{enumerate}{1}
\setlist[subquestion,1]{label=(\alph*)}

\hypersetup{
	citecolor=red,
	colorlinks=true,
	filecolor=red,
	linkcolor=blue,
	linktocpage=true,
	urlcolor=blue
}

{\makeatletter\g@addto@macro\bfseries{\boldmath}\makeatother}
\numberwithin{equation}{section}

\newcommand{\zbar}{\raisebox{0.2ex}{--}\kern-0.6em Z}

\def\td{\textrm{d}}

\def\cP{\mathcal{P}}
\def\cK{\mathcal{K}}

\def\bP{\mathbb{P}}
\def\bK{\mathbb{K}}
\begin{document}

	\title{Entanglement in $\text{T}\bar{\text{T}}$ and root-$\text{T}\bar{\text{T}}$ deformed AdS$_3$/CFT$_2$}

\author{
	Saikat Biswas\thanks{\noindent E-mail:~\texttt{saikatb21@iitk.ac.in}}
}

\affil{
	Department of Physics\\
	Indian Institute of Technology\\
	Kanpur 208016, India
}
	\date{}
	
	\maketitle
	
	\thispagestyle{empty}

	\begin{abstract}
		
		\bigskip
		
		\noindent
		In this work, we investigate the effects of	$\text{T}\bar{\text{T}}$
		and root-$\text{T}\bar{\text{T}}$ deformations on reflected and entanglement entropy in the context of both pure and mixed state entanglement measures. Utilizing a mixed boundary condition framework, we analyze how these deformations modify entanglement structures and explore their implications in three-dimensional AdS space. Our results provide insights into the interplay between solvable irrelevant deformations and quantum information-theoretic quantities, shedding light on the entanglement structure of deformed theories.

	\end{abstract}

	\clearpage
	\tableofcontents
	\clearpage
	\section{Introduction}
	Two-dimensional conformal field theories play central role in theoretical physics due to their exact solvability and infinite-dimensional conformal symmetry, which provide analytic control over strongly coupled and non-perturbative dynamics. The AdS/CFT correspondence \cite{Maldacena:1997re,Gubser:1998bc} offers a precise realization of holography, relating a boundary conformal field theory to a dual gravitational theory in asymptotically AdS spacetime. This duality provides a geometric description of quantum field-theoretic phenomena and serves as a fundamental framework for the study of quantum gravity.

 On the other side, entanglement entropy has proven to be a central quantity in characterizing quantum correlations in bipartite pure states and has played a crucial role in addressing foundational problems in quantum physics. In CFT$_2$, entanglement entropy for bipartite systems is computed using the replica technique \cite{Calabrese:2009qy,Calabrese:2009ez}. On the holographic side, Ryu and Takayanagi proposed a geometric prescription to compute entanglement entropy for a boundary subsystem by evaluating the area of a codimension-two minimal surface in the bulk AdS geometry \cite{Ryu:2006bv,Ryu:2006ef}. This proposal was later generalized covariantly by Hubeny, Rangamani, and Takayanagi \cite{Hubeny:2007xt}, with further justification and refinements provided in \cite{Fursaev:2006ih,Headrick:2010zt,Casini:2011kv,Lewkowycz:2013nqa,Dong:2016hjy,Faulkner:2013ana}.
 
 For mixed quantum states, entanglement entropy does not capture only quantum correlations since it also includes classical ones. To overcome this, various alternative entanglement measures have been introduced, such as entanglement negativity \cite{Vidal:2002zz,Plenio:2005cwa}, entanglement of purification \cite{Takayanagi:2017knl,Caputa:2018xuf}, balanced partial entanglement \cite{Wen:2021qgx,Camargo:2022mme}, odd entanglement entropy \cite{Tamaoka:2018ned}, and reflected entropy \cite{Dutta:2019gen,Jeong:2019xdr}. In CFT$_2$, reflected entropy for bipartite mixed states can be constructed by the replica method \cite{Dutta:2019gen,Jeong:2019xdr}. On the holographic side in AdS$_3$, it has been shown that reflected entropy equals twice the entanglement wedge cross section. This offers a geometric way to characterize correlations in mixed states.

 The $\mathrm{T}\bar{\mathrm{T}}$ deformation, introduced by Zamolodchikov \cite{Zamolodchikov:2004ce}, is an irrelevant deformation of two-dimensional conformal field theories. It is generated by the determinant of the stress-energy tensor. Despite its irrelevance, the deformed theory is remarkably solvable. It admits exact energy spectra and partition functions, and it preserves integrability \cite{Zamolodchikov:2004ce,Cavaglia:2016oda,Smirnov:2016lqw}. These theories show nonlocal behavior in the ultraviolet and allow infinitely many renormalization group flows that converge to the same fixed point. A holographic dual description was proposed as AdS gravity with the boundary at a finite radial cutoff \cite{McGough:2016lol}. This duality has motivated studies of thermodynamics, correlation functions, chaos, and entanglement in the deformed theory \cite{Asrat:2017tzd,Shyam:2017znq,Kraus:2018xrn,Cottrell:2018skz,Taylor:2018xcy,Hartman:2018tkw,Shyam:2018sro,Caputa:2019pam,Giveon:2017myj,Tian:2023fgf,Dei:2024sct}. The impact of the deformation on entanglement measures for both pure and mixed states has been widely explored \cite{Chen:2018eqk,Donnelly:2018bef,Lewkowycz:2019xse,Banerjee:2019ewu,Jeong:2019ylz,Murdia:2019fax,He:2019vzf,Asrat:2019end,Basu:2023bov,Basu:2023aqz,Basu:2024bal,Basu:2024enr,Basu:2025fsf,Chakraborty:2020xwo}. An alternative bulk realization using Dirichlet-Neumann mixed boundary conditions (MBC) was proposed in \cite{Guica:2019nzm}. This provides a complementary view beyond the cutoff AdS proposal and has been applied to study corrections to pure-state entanglement \cite{Banerjee:2024wtl}. Recent studies further expand these analyses to various observables and setups \cite{Apolo:2023aho,Apolo:2023vnm,Apolo:2023ckr,Deng:2020ent,Basu:2024xjq,Basu:2025exh,Hirano:2025alr,Hirano:2025tkq,Hirano:2025cjg,Dey:2025qms}.
 
 The root-$\mathrm{T}\bar{\mathrm{T}}$ deformation is another one-parameter irrelevant deformation of two-dimensional quantum field theories generated by a non-analytic operator. In conformal field theories, this operator is defined as the root-$\mathrm{T}\bar{\mathrm{T}}$ operator and generates a flow that commutes with the standard $\mathrm{T}\bar{\mathrm{T}}$ deformation \cite{Ferko:2022cix}. This structure suggests a rich integrable framework closely related to, yet distinct from, the conventional $\mathrm{T}\bar{\mathrm{T}}$ flow. 
 Motivated by the success of holographic constructions for $\mathrm{T}\bar{\mathrm{T}}$, recent studies have begun exploring entanglement properties of root-$\mathrm{T}\bar{\mathrm{T}}$-deformed thermal CFT$_2$, including pure-state bipartite entanglement for systems with zero and non-zero conserved charges. These analyses involve computing deformed geodesic lengths in the corresponding deformed geometries and comparing the resulting corrections, aiming to shed light on the possible bulk interpretation of the root-$\mathrm{T}\bar{\mathrm{T}}$ deformation.
 
 Motivated by recent developments in the mixed boundary condition description of the bulk dual to $\mathrm{T}\bar{\mathrm{T}}$-deformed theories, the authors of \cite{Banerjee:2024wtl} evaluated the entanglement entropy in $\mathrm{T}\bar{\mathrm{T}}$-deformed finite-temperature CFT$_2$ with both zero and non-zero conserved charges, where the bulk computations were performed using the MBC prescription. While this framework has been shown to successfully reproduce pure-state entanglement entropy, a natural and important question arises as to whether the mixed boundary condition approach remains valid for mixed-state entanglement measures. In this work, we address this issue by computing the correction to the entanglement wedge cross section, which is holographically related to the reflected entropy, and by explicitly comparing our results with those obtained using the AdS cutoff prescription and the CFT result available in the existing literature \cite{Basu:2024bal,Basu:2024enr}. The agreement between the two approaches provides a nontrivial consistency check of the mixed boundary condition framework for mixed-state entanglement measures.
 
 Having established the applicability of the mixed boundary condition prescription to mixed-state entanglement in $\mathrm{T}\bar{\mathrm{T}}$-deformed theories, a further intriguing question is whether this framework can be consistently extended to unconventional irrelevant deformations, such as the root-$\mathrm{T}\bar{\mathrm{T}}$ deformed of CFT$_2$. We investigate this question by exploring the pure-state entanglement structure of root-$\mathrm{T}\bar{\mathrm{T}}$-deformed finite-temperature CFT$_2$ with and without a conserved charge. In contrast to previous approaches based on the AdS cutoff prescription, our bulk computations employ the mixed boundary condition description, thereby testing its robustness and flexibility in a broader class of deformed AdS/CFT setups.

    In this article, we divide our objective into two parts. In the first part, we compute the EWCS in the $\text{T}\bar{\text{T}}$ deformed BTZ and rotating BTZ geometries using the mixed boundary condition. Later, by expanding the result with respect to the deformation parameter, we obtain the first-order correction. We can then see that these results match the field-theoretic computations available in the literature, as well as the result computed using the AdS cut-off prescription. In the next part, we use the partition function for a single interval on the field theory side, and applying the root-$\text{T}\bar{\text{T}}$ deformation, we can obtain the correction formula for an $n$-sheeted manifold. Further computation yields the correction for both zero and finite conserved charge thermal CFT$_2$. Using the deformed geometry, we again compute the geodesic length and the first-order correction, which matches the field theory result. Further, we analyze QNEC for entanglement entropy, timelike and spacelike.\footnote{The QNEC is rigorously established for local QFTs while the $T\bar T$–deformed theory is non-local at scales $\lesssim \sqrt{\mu}$, we restrict our analysis to intervals $l \gg \sqrt{\mu}$ where the theory can be treated as effectively quasi-local.}
  
  Further, our article is organised as follows. In \cref{sec2}, we briefly review the holographic description of the $\text{T}\bar{\text{T}}$ deformation, the root-$\text{T}\bar{\text{T}}$ deformation, and the holographic formulation of reflected entropy. In \cref{sec3}, we compute the entanglement wedge cross section for various boosted subsystems in geometries with both vanishing and non-vanishing angular momentum. In \cref{sec4}, we evaluate the corrections to the entanglement entropy for a boosted single interval in a two-dimensional finite-temperature CFT with zero and finite conserved charge arising from the root-$\text{T}\bar{\text{T}}$ deformation. In \cref{sec5}, we study the holographic entanglement entropy in the root-$\text{T}\bar{\text{T}}$ deformed theory for both zero and finite chemical potential, and analyse the resulting bounds on the chemical potential $\mu$ based on the HEE. In \cref{sec6}, we investigate the Quantum Null Energy Condition. Finally, we summarise our results and conclude in \cref{sec7}.


     \section{Review of earlier literature}\label{sec2}

     In this section, we now briefly review $\text{T}\bar{\text{T}}$ and root-$\text{T}\bar{\text{T}}$ deformed CFT$_2$s, and their corresponding bulk duals in terms of the mixed boundary condition. Subsequently, we discuss the mixed state measure known as the reflected entropy and its holographic dual.

    
	\subsection{Holographic description of $\text{T}\bar{\text{T}}$}

    We begin by reviewing the $\text{T}\bar{\text{T}}$ deformation of a CFT$_2$, and the effects on the underlying metric $\gamma^{\alpha\beta}$ and the field theory stress energy tensor $T_{\alpha\beta}$ due to this deformation \cite{Guica:2019nzm}.
    The $\text{T}\bar{\text{T}}$ deformation is a universal irrelevant deformation applicable to any two-dimensional local conformal field theory, defined by the flow equation  
	\begin{gather}\label{S_mu}
		\partial_{\mu}S^{[\mu]} = -\frac{1}{2}\int\dd^2x\sqrt{\gamma}\left(T^{\alpha \beta}T_{\alpha \beta} - T^2\right)_\mu\,.
	\end{gather}
	 Here $S^{[\mu]}$ is the boundary action in the deformed theory, and we adopt the signature convention of \cite{Guica:2019nzm}. Now, this deformation generates an integrable flow in the space of field theories. The stress tensor is computed from the variation of the deformed action for the metric. This formulation is valid in the regime where physical scales are much larger than $\sqrt{\mu}$; beyond this, non-local effects dominate, and a local field-theoretic description fails.
	
	Due to its double-trace nature and source-dependent deformation, the standard holographic dictionary must be adjusted. These deformations often result in mixed boundary conditions, and for large-$N$ CFTs, a deformed holographic dictionary can be systematically derived. Varying Eq.~\eqref{S_mu} with respect to the metric gives
	\begin{gather}
		\delta S^{[\mu]} = \left(\frac{1}{2}\int\dd^2x\sqrt{\gamma}T_{\alpha\beta}\delta^{\alpha\beta}\right)^{[\mu]},
	\end{gather}
	leading to the flow equations:
	\begin{gather}
		\partial_{\mu}\gamma^{\alpha\beta} = 2\left(T^{\alpha\beta} - \gamma^{\alpha\beta}T\right), \label{flow_eqn1} \\
		\partial_{\mu}\left(\sqrt{\gamma}T_{\alpha\beta}\right) = \sqrt{\gamma}\left(2TT_{\alpha\beta} - 2T_{\alpha\eta}T^{\eta}_\beta - \frac{1}{2}\gamma_{\alpha\beta}\mathcal{O}_{\text{T}\bar{\text{T}}}\right), \label{flow_eqn2}
	\end{gather}
	where $\mathcal{O}_{\text{T}\bar{\text{T}}} = T^{\alpha\beta}T_{\alpha\beta} - T^2$. Introducing $\hat{T}_{\alpha\beta} = T_{\alpha\beta} - \gamma_{\alpha\beta}T$, we rewrite the above as:
	\begin{gather}
		\partial_{\mu}\gamma_{\alpha\beta} = -2\hat{T}_{\alpha\beta}, \label{flow_eqn3} \\
		\partial_{\mu}\hat{T}_{\alpha\beta} = -\hat{T}_{\alpha\eta}\hat{T}^\eta_\beta. \label{flow_eqn4}
	\end{gather}
	
	Using the relations $\partial_{\mu}\sqrt{\gamma} = \sqrt{\gamma}T$ and $\partial_{\mu}T = -T^{\alpha\beta}T_{\alpha\beta}$, further differentiation yields the following
	\begin{gather}
		\partial_{\mu}^3\gamma_{\alpha\beta} = 0, \label{flow-eqn5} \\
		\partial_{\mu}\hat{T}_{\alpha\beta} = -\hat{T}_{\alpha\eta}\hat{T}^\eta_\beta. \label{flow-eqn6}
	\end{gather}
	
	Since $\hat{T}_{\alpha\eta}\hat{T}^\eta_\beta$ is $\mu$ independent, we can solve \eqref{flow-eqn5} and \eqref{flow-eqn6} to obtain the deformed metric and stress tensor expectation values.
	\begin{gather}
		\gamma_{\alpha\beta}^{[\mu]} = \gamma_{\alpha\beta}^{[0]} - 2\mu\hat{T}_{\alpha\beta}^{[0]} + \mu^2 \hat{T}_{\alpha\rho}^{[0]}\gamma^{[0]\rho\sigma}\hat{T}_{\sigma\beta}^{[0]}, \\
		\hat{T}_{\alpha\beta}^{[\mu]} = \hat{T}_{\alpha\beta}^{[0]} - \mu\hat{T}_{\alpha\rho}^{[0]}\gamma^{[0]\rho\sigma}\hat{T}_{\sigma\beta}^{[0]}.
	\end{gather}
	
	For pure gravity, we identify:
	\begin{gather}
		\gamma_{\alpha\beta}^{[0]} = g_{\alpha\beta}^{[0]}, \quad \hat{T}_{\alpha\beta}^{[0]} = \frac{1}{8 \pi G l_{\text{AdS}}}g_{\alpha\beta}^{[2]}, \quad \hat{T}_{\alpha\rho}^{[0]}\gamma^{[0]\rho\sigma}\hat{T}_{\sigma\beta}^{[0]} = \frac{1}{(4\pi G l_{\text{AdS}})^2}g_{\alpha\beta}^{[4]}.
	\end{gather}
	
	The Fefferman-Graham (FG) expansion of the AdS$_3$ metric is:
	\begin{gather}
		\dd s^2 = l_{\text{AdS}}^2\frac{\dd \rho^2}{4 \rho^2} + \frac{1}{\rho}g_{\alpha\beta}(x, \rho) \dd x^\alpha \dd x^\beta, 
        \end{gather}
	with 
        \begin{gather}
        g_{\alpha\beta}(x, \rho) = g_{\alpha\beta}^{[0]} + \rho g_{\alpha\beta}^{[2]} + \rho^2 g_{\alpha\beta}^{[4]} + \dots
	\end{gather}
	
	Imposing Dirichlet boundary conditions at a finite radial slice $\rho = \rho_c$, the deformed stress tensor is identified as:
	\begin{align}
		T_{\alpha\beta}^{[\mu]} = T_{\alpha\beta}^{\text{BY}}(\rho_c) - \frac{g_{\alpha\beta}(\rho_c)}{8\pi G l_{\text{AdS}}},
	\end{align}
	where $T_{\alpha\beta}^{\text{BY}}(\rho_c)$ is the Brown-York stress tensor and $g_{\alpha\beta}(\rho_c)$ is the induced metric at $\rho = \rho_c$.  One can see \cite{Banerjee:2024wtl} for further discussion.
	\subsection{Deformed Spacetime Geometry}
	In this part, we start with the most general solution of Einstein's equations with $\Lambda = -2/l^2_{\text{AdS}}$ in FG form, which is as follows
	\begin{align}\label{FG_metric}
		\dd s^2 = l^2_{\text{AdS}}\frac{\dd \rho^2}{4 \rho^2} + \frac{\dd u \dd v}{\rho} + 	\left(L(u)\dd u^2 + \bar{L}(v)\dd v^2\right) + \rho L(u)\bar{L}(v) \dd u \dd v.
	\end{align}
	Now on a constant slice $\rho = \rho_c$, the metric becomes
	\begin{align}
		\rho_c \dd s^2 = \left(\dd u + \rho_c \bar{L}(v) \dd v\right)\left(\dd v + \rho_c L(u) \dd u\right).
	\end{align}
	Further this metric can be mapped to a flat form  $\dd s^2 = \dd U \dd V$ via
	\begin{gather}
		U = u + \rho_c \int^v \bar{L}(v') \dd v', \quad V = v + \rho_c \int^u L(u') \dd u'\, ,
	\end{gather}
	and assuming constant $L$ and $\bar{L}$, the inverse map becomes
	\begin{gather}\label{map_1}
		u = \frac{U - \rho_c \bar{L} V}{1 - \rho_c^2 L \bar{L}}, \quad v = \frac{V - \rho_c L U}{1 - \rho_c^2 L \bar{L}}.
	\end{gather}
	Next we can substituting \eqref{map_1} into \eqref{FG_metric} which yields the deformed metric
	\begin{align}\label{deff_metric_1}
		\dd s^2_{\text{def}} = l^2_{\text{AdS}}\frac{\dd \rho^2}{4\rho^2} 
		+ \frac{\left(\dd s^{[0]}_{\text{def}}\right)^2}{\rho} 
		+ \left(\dd s^{[2]}_{\text{def}}\right)^2 
		+ \rho \left(\dd s^{[4]}_{\text{def}}\right)^2,
	\end{align}
	where the components of the deformed metric are given as
	\begin{align}
		\left(\dd s^{[0]}_{\text{def}}\right)^2 &= 
		\frac{(\dd U - \rho_c \bar{L}\dd V)(\dd V - \rho_c L\dd U)}{(1 - \rho_c^2 L\bar{L})^2}, \notag\\
		\left(\dd s^{[2]}_{\text{def}}\right)^2 &= 
		\frac{(1 + \rho_c^2 L\bar{L})(L\dd U^2 + \bar{L}\dd V^2) - 4\rho_c L\bar{L}\dd U\dd V}{(1 - \rho_c^2 L\bar{L})^2}, \notag\\
		\left(\dd s^{[4]}_{\text{def}}\right)^2 &= L\bar{L}\left(\dd s^{[0]}_{\text{def}}\right)^2. \notag
	\end{align}
	Finally, to match the mixed boundary condition (MBC) interpretation, we identify
	\begin{align}
		\rho_c = -\frac{\mu}{4\pi G l_{\text{AdS}}} = -2\mu\, .
	\end{align}
    Here we can use this expression in the deformed metric \cref{deff_metric_1}, and express it in terms of $\mu$, where the deformation parameter is allowed to have both signs.	\subsection{Root-$\text{T}\bar{\text{T}}$ Deformation}
	
	The root-$\text{T}\bar{\text{T}}$ deformation presents a subtle and non-trivial extension of irrelevant deformations, particularly from a holographic perspective. It belongs to a class of deformations characterized by the property that the combination \cite{Ebert:2023tih}
	\begin{align} \label{root1}
		\int d^2x \sqrt{\gamma^{(\mu)}}\, T^{(\mu)}_{\alpha\beta}\, \delta \gamma^{(\mu)\alpha\beta}\, ,
	\end{align}
	which remains invariant under variations of the deformation parameter $\mu$. This class includes trivial deformations such as diffeomorphisms and Weyl rescalings,
     which do not modify the physical content of the theory.	Though the root-$\text{T}\bar{\text{T}}$ deformation is distinct from such trivial transformations. Multiple pieces of evidence in the literature suggest that it represents a genuinely new type of deformation, with non-trivial implications for the structure of the theory. In what follows, we aim to distinguish the root-$\text{T}\bar{\text{T}}$ deformation from other deformations satisfying \cref{root1} and outline a general framework for understanding its consequences. To this end, we enumerate a set of desirable properties that the deformed theory should obey and construct the most general form of the boundary metric $\gamma_{\alpha\beta}^{(\mu)}$ and stress-energy tensor $T_{\alpha\beta}^{(\mu)}$ consistent with these principles. This leads us to a well-defined proposal for the finite-volume spectrum of a root-$\text{T}\bar{\text{T}}$ deformed conformal field theory defined on a cylinder.
	
	\vspace{1em}
	\noindent The deformation is required to satisfy the following conditions:
	
	\begin{enumerate}[label=(\roman*)]
		\item \textbf{Tracelessness of the Stress Tensor:} Both the undeformed and deformed stress-energy tensors are traceless with respect to their respective metrics:
		\begin{align}
			\gamma^{(0)\alpha\beta} T^{(0)}_{\alpha\beta} = 0\,, \qquad
			\gamma^{(\mu)\alpha\beta} T^{(\mu)}_{\alpha\beta} = 0\,.
		\end{align}
		
		\item \textbf{Group Structure of the Deformation:} The deformation must form a one-parameter group under composition:
		\begin{align}
			&\left(\gamma_{\alpha\beta}^{(0)},\, T_{\alpha\beta}^{(0)}\right)
			\xrightarrow{\mu_1}
			\left(\gamma_{\alpha\beta}^{(\mu_1)},\, T_{\alpha\beta}^{(\mu_1)}\right)
			\xrightarrow{\mu_2}
			\left(\gamma_{\alpha\beta}^{(\mu_1 + \mu_2)},\, T_{\alpha\beta}^{(\mu_1 + \mu_2)}\right)\,, \\
			&\left(\gamma_{\alpha\beta}^{(0)},\, T_{\alpha\beta}^{(0)}\right)
			\xrightarrow{\mu_1 + \mu_2}
			\left(\gamma_{\alpha\beta}^{(\mu_1 + \mu_2)},\, T_{\alpha\beta}^{(\mu_1 + \mu_2)}\right)\,.
		\end{align}
		
		\item \textbf{Commutativity with $\text{T}\bar{\text{T}}$ Deformation:} The root-$\text{T}\bar{\text{T}}$ deformation must commute with the ordinary $\text{T}\bar{\text{T}}$ deformation.
	\end{enumerate}
	
	\vspace{1em}
	\noindent Assuming a conformal seed theory with a traceless stress tensor, the only available Lorentz-invariant scalar quantity is:
	\begin{align}
		\mathcal{R}^{(0)} = \sqrt{\frac{1}{2} T^{(0)\alpha\beta} T^{(0)}_{\alpha\beta}}\,.
	\end{align}
	
	Another tensor structure can be built from:
	\begin{align}
		\left(T^2\right)_{\alpha\beta} = T^{(0)}_{\alpha\gamma} T^{(0)\gamma}_\beta\,.
	\end{align}
	But due to the tracelessness condition, this reduces to:
	\begin{align}
		\left(T^2\right)_{\alpha\beta} = \left(\mathcal{R}^{(0)}\right)^2 \gamma^{(0)}_{\alpha\beta}\,,
	\end{align}
	implying that higher powers of the stress tensor also collapse to linear combinations of $\gamma^{(0)}_{\alpha\beta}$ and $T^{(0)}_{\alpha\beta}$, with $\mathcal{R}^{(0)}$-dependent coefficients. Consequently, no new independent tensor structures emerge.
	
	\vspace{1em}
	\noindent Based on these considerations, the most general ansatz for the deformed metric and stress-energy tensor is:
	\begin{align}
		\gamma^{(\mu)}_{\alpha\beta} &= f_1(\mu)\, \gamma^{(0)}_{\alpha\beta} + \frac{f_2(\mu)}{\mathcal{R}^{(0)}}\, T^{(0)}_{\alpha\beta}\,, \\
		T^{(\mu)}_{\alpha\beta} &= f_3(\mu)\, T^{(0)}_{\alpha\beta} + f_4(\mu)\, \mathcal{R}^{(0)}\, \gamma^{(0)}_{\alpha\beta}\,,
	\end{align}
	where the functions $f_i(\mu)$ are determined by enforcing the conditions listed above. As shown in \cite{Ebert:2023tih}, the unique solution that satisfies all the requirements is given by:
	\begin{align} \label{root_ttbar_final}
		\gamma^{(\mu)}_{\alpha\beta} &= \cosh(\mu)\, \gamma^{(0)}_{\alpha\beta} + \frac{\sinh(\mu)}{\mathcal{R}^{(0)}}\, T^{(0)}_{\alpha\beta}\,, \\
		T^{(\mu)}_{\alpha\beta} &= \cosh(\mu)\, T^{(0)}_{\alpha\beta} + \sinh(\mu)\, \mathcal{R}^{(0)}\, \gamma^{(0)}_{\alpha\beta}\,.
	\end{align}
	
	These expressions define the flow of the boundary data under the root-$\text{T}\bar{\text{T}}$ deformation. Similar forms can also be derived for non-conformal seed theories, though the analysis may involve additional structures due to the presence of a non-vanishing trace for the stress tensor.
	\subsection{The Deformed Spacetime for root-$\text{T}\bar{\text{T}}$}\label{rootttbar}
	
	In the case of the root-$\text{T}\bar{\text{T}}$, the boundary metric and the stress-energy tensor take the form given in \cref{root_ttbar_final}. To analyze this deformation in the bulk, we begin by considering the Fefferman–Graham expansion of the Bañados geometry. In this context, the FG components are given as
	\begin{align}
		g_{\alpha\beta}^{(0)} dx^\alpha dx^\beta &= du\,dv\,, \\
		g_{\alpha\beta}^{(2)} dx^\alpha dx^\beta &= \mathcal{L}(u)\, du^2 + \bar{\mathcal{L}}(v)\, dv^2\,, \\
		g_{\alpha\beta}^{(4)} dx^\alpha dx^\beta &= \mathcal{L}(u)\bar{\mathcal{L}}(v)\, du\, dv\,.
	\end{align}
	
	Using these components, the deformed boundary metric $\gamma^{(\mu)}_{\alpha\beta}$ derived from the root-$\text{T}\bar{\text{T}}$ deformation takes the form
	\begin{align}
		\gamma_{\alpha\beta}^{(\mu)} = \cosh(\mu)\, g_{\alpha\beta}^{(0)} + \frac{\sinh(\mu)}{2\sqrt{\mathcal{L}(u)\bar{\mathcal{L}}(v)}}\, g_{\alpha\beta}^{(2)}\,.
	\end{align}
	
	Similarly, the deformed boundary stress-energy tensor becomes:
	\begin{align}
		\tilde{T}_{\alpha\beta} = \frac{1}{2} \cosh(\mu)\, g_{\alpha\beta}^{(2)} + \sqrt{\mathcal{L}(u)\bar{\mathcal{L}}(v)}\, \sinh(\mu)\, g_{\alpha\beta}^{(0)}\,.
	\end{align}
	
	Here, we have adopted the convention $4\pi G l = 1$ and used the relations from the FG expansion, particularly $g_{\alpha\beta}^{(0)} = 2 T_{\alpha\beta}$, to compute the scalar quantity $\mathcal{R}$.
	
	Next, we perform a coordinate transformation to bring the deformed metric into a canonical form. Define the new lightcone coordinates $(U, V)$ as follows:
	\begin{align} \label{root_UV}
		dU &= \cosh\left(\frac{\mu}{2}\right) du + \sqrt{\frac{\bar{\mathcal{L}}(v)}{\mathcal{L}(u)}}\, \sinh\left(\frac{\mu}{2}\right) dv\,, \notag\\
		dV &= \sqrt{\frac{\bar{\mathcal{L}}(v)}{\mathcal{L}(u)}}\, \sinh\left(\frac{\mu}{2}\right) du + \cosh\left(\frac{\mu}{2}\right) dv\,.
	\end{align}
	
	This transformation has the elegant property that it brings the deformed metric back to the flat form:
	\begin{align}
		\gamma_{\alpha\beta}^{(\mu)} dx^\alpha dx^\beta = dU\, dV\,.
	\end{align}
	
	Thus, under this change of coordinates, the root-$\text{T}\bar{\text{T}}$ deformed boundary geometry retains the flat structure in the new frame, albeit with non-trivial dependence encoded in the coordinate transformation itself.
	\subsection{Reflected Entropy and Its Holographic Dual}
	Reflected entropy is a well-established measure of correlations for mixed quantum states, originally proposed in \cite{Dutta:2019gen,Jeong:2019xdr}. 
	From a field-theoretic perspective, it provides a natural extension of entanglement entropy beyond pure states and captures both classical and quantum correlations between two subsystems.
	Given a mixed density matrix $\rho_{AB}$,  the reflected entropy $S_R(A:B)$ is defined by first constructing its canonical purification and then computing the entanglement entropy across a reflection-symmetric bipartition of the purified Hilbert space.
	
	In holographic theories, reflected entropy admits a remarkably simple geometric dual. 
	It is given by twice the entanglement wedge cross section  \cite{Dutta:2019gen,Jeong:2019xdr},
	\begin{align}
		S_R(A:B) = 2\,E_W(A:B),
	\end{align}
	where $E_W(A:B)$ denotes the minimal-area surface of codimension one that bisects the entanglement wedge $\mathcal{E}_{AB}$ associated with the reduced density matrix $\rho_{AB}$.
	The entanglement wedge is defined as the bulk region bounded by the boundary subregion $A\cup B$ and the extremal surface(s) $\Gamma_{AB}$ homologous to it \cite{Takayanagi:2017knl,Nguyen:2017yqw}.
	
	In the AdS$_3$/CFT$_2$ correspondence, these extremal surfaces reduce to geodesics.
	An efficient method for evaluating the EWCS employs the embedding of AdS$_3$ into the flat space $\mathbb{R}^{2,2}$,
	\begin{align}
		\td s^2 = \eta_{\mu\nu}\,\td X^\mu \td X^\nu,
		\qquad
		X^\mu X_\mu = -1,
		\qquad
		\mu = 1,\ldots,4\,,
		\label{AdS-embedding}
	\end{align}
	where conformally invariant distances are encoded in the scalar products $\zeta_{ij} = -X_i \cdot X_j$.
	
	Within this framework, the EWCS for two disjoint intervals
	$A=[X_1,X_2]$ and $B=[X_3,X_4]$ is given by \cite{Kusuki:2019evw,Kusuki:2019rbk}
	\begin{align}
		E_W(A:B)
		= \frac{1}{4G_N}
		\cosh^{-1}\!\left(\frac{1+\sqrt{u}}{\sqrt{v}}\right),
		\label{EW-disj-formula}
	\end{align}
	with
	\begin{align}
		u = \frac{\zeta_{12}\zeta_{34}}{\zeta_{13}\zeta_{24}},
		\qquad
		v = \frac{\zeta_{14}\zeta_{23}}{\zeta_{13}\zeta_{24}}.
	\end{align}
	For two adjacent intervals $A=[X_1,X_2]$ and $B=[X_2,X_3]$, the EWCS simplifies to \cite{Basu:2023jtf}
	\begin{align}
		E_W(A:B)
		= \frac{1}{4G_N}
		\cosh^{-1}\!\left(
		\sqrt{\frac{2\,\zeta_{12}\zeta_{23}}{\zeta_{13}}}
		\right).
		\label{EW-adj-formula}
	\end{align}
	
	The introduction of reflected entropy is thus motivated by the existence of a sharp geometric dual in holographic theories, extending the Ryu--Takayanagi prescription to mixed states and providing a refined probe of correlations beyond entanglement entropy.

   
	\section{EWCS for $\text{T}\bar{\text{T}}$ }\label{sec3}

Having reviewed the basics, in this section we investigate the validity of the MBC formalism in $\text{T}\bar{\text{T}}$ deformed CFT$_2$s for mixed state entanglement measures using the reflected entropy. The MBC formalism in $\text{T}\bar{\text{T}}$ deformed CFT$_2$s has already been verified for the pure state entanglement measures. To this end, we determine the first order correction to the entanglement wedge cross section (EWCS) between two disjoint, two adjacent, and a single subsystem on the asymptotic boundary of BTZ black holes with both vanishing and finite chemical potential.\footnote{The field theory computation of the reflected entropy in $\text{T}\bar{\text{T}}$ are available in \cite{}.} We demonstrate that the results obtained using the MBC formalism align with those available in the literature.

	\subsection{Vanishing Chemical Potential}
	
	In this section, we discuss the $\text{T}\bar{\text{T}}$ deformed BTZ black hole with a vanishing chemical potential described in \cite{Banerjee:2024wtl}. Setting the chemical potential $\Omega$ to zero allows us to set the chiral functions $L = \bar{L}$ in \cref{FG_metric}, and using \cref{deff_metric_1} we have the deformed bulk geometry as
	\begin{gather}
		ds^2 = \frac{d\rho^2}{4 \rho^2} + \frac{L (\rho + 2\mu)(1 + 2\mu L^2 \rho)}{\rho(1 - 4\mu^2 L^2)^2}(dU^2 + dV^2) + \frac{1 + 4\mu^2 L^2 + L^2 \rho (\rho + 8\mu + 4\mu^2 L^2 \rho)}{\rho (1 - 4\mu^2 L^2)^2} dU dV\, .
	\end{gather}
Using the coordinate transformations 
        \begin{gather}\label{BTZ trans}
		\rho = \frac{r^2 - 2L - r\sqrt{r^2 - 4L}}{2L^2}, \quad U = X + T, \quad V = X - T\, .
	\end{gather}
    the above deformed geometry may be expressed as 
    \begin{gather}
		ds^{2}
		= \frac{dr^{2}}{\,r^{2}-4L\,}
		- (r^{2}-4L)\,\frac{dT^{2}}{(1+2\mu L)^{2}}
		+ r^{2}\,\frac{dX^{2}}{(1-2\mu L)^{2}}\, ,
		\label{BTZ_metric}
	\end{gather}
which reduces to the well known BTZ black hole with horizon radius $r_h = 2 \sqrt{L}$ for $\mu \to 0$. This deformed geometry may be embedded in $\mathbb{R}^{2,2}$ using the following embedding coordinates
	\begin{align}
		X_{1} &= \frac{r}{2\sqrt{L}} \cosh\!\left(2a(\mu)\sqrt{L}\,X\right),
		&
		X_{2} &= \frac{r}{2\sqrt{L}} \cosh\!\left(2b(\mu)\sqrt{L}\,T\right),\notag
		\\
		X_{3} &= \frac{r}{2\sqrt{L}} \sinh\!\left(2a(\mu)\sqrt{L}\,X\right),
		&
		X_{4} &= \frac{r}{2\sqrt{L}} \sinh\!\left(2b(\mu)\sqrt{L}\,T\right),
		\label{Em_BTZ}
	\end{align}
    where
	\begin{align}
		a(\mu) = \frac{1}{1 - 2\mu L},
		\qquad
		b(\mu) = \frac{1}{1 + 2\mu L}.
	\end{align}
    From \cref{BTZ_metric} we may further obtain the relation between the horizon radius $r_h$ and the inverse temperature $\beta$ and the deformation parameter $\mu$ as 
    \begin{align}
		r_h =2 \sqrt{L}=
		\frac{\beta - \sqrt{\beta^{2} - 8\pi^{2}\mu}}
		{2\pi \mu}.
		\label{L_beta}
	\end{align}
This highlights the fact that in contrast to the conventional cutoff description where the $\text{T}\bar{\text{T}}$ deformation manifests as a radial cutoff at some finite $r=r_c$, in the alternate MBC formalism the deformation leads to a modification of the radius of the BTZ black hole.

	\subsubsection{Two disjoint intervals}
	Now, we consider two disjoint boosted intervals
$A=\left[\left(x_1,t_1\right),\left(x_2,t_2\right)\right]$ and
$B=\left[\left(x_3,t_3\right),\left(x_4,t_4\right)\right]$,
 in a finite-temperature
$\text{T}\bar{\text{T}}$-deformed CFT$_2$ without conserved charge.
The bulk dual is a planar BTZ black hole,
the entanglement wedge cross-section, which is dual to the reflected
entropy is computed using the embedding formalism in the MBC prescription, following \cref{Em_BTZ,EW-disj-formula}.
 Then, replacing $L$ according to \cref{L_beta}, we obtain the expression for the entanglement wedge cross section in the deformed theory in the non-perturbative regime. However, the non-perturbative result currently lacks a consistency check from the field-theoretic perspective. Therefore, we expand the result perturbatively in the deformation parameter $\mu$. The leading-order and first-order corrections arising due to the $\text{T}\bar{\text{T}}$ deformation are given by
	\begin{align}
		E_W\left(A:B\right)=&\frac{1}{4G_N}\cosh^{-1}\!\left[\frac{1+\sqrt{\eta\bar{\eta}}}{\sqrt{\left(1-\eta\right)\left(1-\bar{\eta}\right)}}\right]\notag\\
		&+\frac{\mu\pi^3\sqrt{\eta\bar{\eta}}}{2G_N\beta^3\left(\sqrt{\eta}+\sqrt{\bar{\eta}}\right)}\left[\bP_{21}+\bP_{43}-\bP_{32}-\bP_{41}\right]\notag\\
		&+\frac{\mu\pi^3}{2G_N\beta^3\left(\sqrt{\eta}+\sqrt{\bar{\eta}}\right)}\left[\bP_{31}+\bP_{42}-\bP_{32}-\bP_{41}\right], \label{EW-disj}
	\end{align}
	where we have defined
	\begin{align}
		\bP_{ij}=\cP_{ij}+\bar\cP_{ij}=x_{ij}\left[\coth\!\left(\frac{\pi\left(x_{ij}+t_{ij}\right)}{\beta}\right)+\coth\!\left(\frac{\pi\left(x_{ij}-t_{ij}\right)}{\beta}\right)\right]\, , \label{bP-ij}
	\end{align}
	and $\eta$ and $\bar{\eta}$ is represented by 
	\begin{align}
		\eta = \frac{\sinh \left(\frac{\pi  \left(x_{21}+t_{21}\right)}{\beta}\right) \sinh \left(\frac{\pi  \left(x_{43}+t_{43}\right)}{\beta}\right)}{\sinh \left(\frac{\pi  \left(x_{31}+t_{31}\right)}{\beta}\right) \sinh \left(\frac{\pi  \left(x_{42}+t_{42}\right)}{\beta}\right)}\,, \qquad \bar{\eta} =\frac{\sinh \left(\frac{\pi  \left(x_{21}-t_{21}\right)}{\beta}\right) \sinh \left(\frac{\pi  \left(x_{43}-t_{43}\right)}{\beta}\right)}{\sinh \left(\frac{\pi  \left(x_{31}-t_{31}\right)}{\beta}\right) \sinh \left(\frac{\pi  \left(x_{42}-t_{42}\right)}{\beta}\right)} \, .
	\end{align}
	This result is in complete agreement with the corresponding CFT computation reported in \cite{Basu:2024bal}. It also matches, up to first order, with the $\text{T}\bar{\text{T}}$-deformed result obtained in \cite{Basu:2024bal} using the cut-off description for a positive deformation parameter.	
	\subsubsection{Two adjacent intervals}
	Next, we examine the configuration of two adjacent boosted intervals, denoted $A = \left[\left(x_1, t_1\right), \left(x_2, t_2\right)\right]$ and $B = \left[\left(x_2, t_2\right), \left(x_3, t_3\right)\right]$. These intervals are defined on a finite-temperature cylinder with no conserved charge, where the bulk AdS geometry is described by a BTZ black hole. To compute the EWCS for this configuration, we use the relation \cref{EW-adj-formula}. In the $\text{T}\bar{\text{T}}$-deformed scenario, we employ the embedding coordinates given in \cref{Embedding-coordinates} within the MBC framework. Subsequently, we replace the deformed horizon identification with the original inverse temperature of the theory using the relation \cref{L_beta}. Expanding this result for a small deformation parameter $\mu$, we express the EWCS as
	\begin{align}
		E_W(A:B)
		&= \frac{1}{8G_N}
		\log\!\left[
		\frac{\beta}{\pi \epsilon_c}
		\frac{
			\sinh\!\left(\frac{\pi (x_{21}+t_{21})}{\beta}\right)
			\sinh\!\left(\frac{\pi (x_{32}+t_{32})}{\beta}\right)
		}{
			\sinh\!\left(\frac{\pi (x_{31}+t_{31})}{\beta}\right)
		}
		\right]
		\nonumber\\
		&\quad
		+ \frac{1}{8G_N}
		\log\!\left[
		\frac{\beta}{\pi \epsilon_c}
		\frac{
			\sinh\!\left(\frac{\pi (x_{21}-t_{21})}{\beta}\right)
			\sinh\!\left(\frac{\pi (x_{32}-t_{32})}{\beta}\right)
		}{
			\sinh\!\left(\frac{\pi (x_{31}-t_{31})}{\beta}\right)
		}
		\right]
		\nonumber\\
		&\quad
		+ \frac{\pi^3 \mu}{2G_N \beta^3}
		\left(\bP_{21} + \bP_{32} - \bP_{31}\right)
		- \frac{\pi^2 \mu}{2G_N \beta^2}.
	\end{align}
	
	In this expression, if we neglect the final $\mu$-dependent term that does not involve the interval configuration, the first term corresponds to the standard undeformed EWCS, while the second term shows exact agreement with the CFT result obtained in \cite{Basu:2024bal} for small values of the deformation parameter. Interestingly, this correction also matches the $\text{T}\bar{\text{T}}$-deformed bulk theory result discussed in \cite{Basu:2024bal}, corresponding to the cut-off description, in the limit of small and positive deformation parameter at high temperature.
	\subsubsection{A single interval}
	Finally, we consider a single interval $A = \left[\left(0, 0\right), \left(\ell, t\right)\right]$ in a thermal state of the deformed CFT$_2$. To compute the reflected entropy holographically, one must introduce two auxiliary intervals, $B_1 = \left[\left(-L, -T\right), \left(0, 0\right)\right]$ and $B_2 = \left[\left(\ell, t\right), \left(L, T\right)\right]$, such that $B = B_1 \cup B_2$ sandwich the interval $A$.	In general, the wedge entanglement cross section $\tilde{E}_W(A:B)$ is the sum of the EWCS between $A$ and $B_1$ and between $A$ and $B_2$
	\begin{align}\label{UP_RE}
		\tilde{E}_W\left(A:A^c\right) = E_W\left(A:B_1\right) + E_W\left(A:B_2\right)\, .
	\end{align}
	To compute the upper bound of the reflected entropy, one considers the bipartite limit where the auxiliary intervals become infinitely extended, i.e., $L \to \infty$. To compute this, we use the adjacent interval result computed earlier and put the corresponding limits and boundary conditions. Threfor the full entanglement wedge cross section becomes
	\begin{align}\label{EW_single}
		\tilde{E}_W\left(A:A^c\right)&=\frac{1}{4G}\log \left(\frac{2 \beta ^2 \left(\cosh \left(\frac{2 \pi  \ell}{\beta }\right)-\cosh \left(\frac{2 \pi  t}{\beta }\right)\right)}{\pi ^2 \epsilon _c^2}\right)\\
		&\quad +\frac{2 \pi ^3 \ell \mu  \sinh \left(\frac{2 \pi  \ell}{\beta }\right)}{G\beta ^3 \left(\cosh \left(\frac{2 \pi  \ell}{\beta }\right)-\cosh \left(\frac{2 \pi  t}{\beta }\right)\right)}\\
		&\quad -\frac{ \pi ^2 \mu }{G\beta ^2}-\frac{2 \pi ^3 \mu  \ell}{G\beta ^3}-\frac{ \pi  \ell}{2G\beta }\, .
	\end{align}
	This expression, if we drop $-\frac{ \pi ^2 \mu }{G\beta ^2}$ for high temperature limit, then the other $\mu$ proportional terms represent the correction to the single interval in this scenario. This shows a consistent match with the field-theoretic result computed in \cite{Basu:2024bal}. It also agrees with cut off theory result for positive and small deformation parameter in the perturbative limit showed in \cite{Basu:2024bal}.
	
	One interesting thing to note at here, that the expression obtained in \cref{EW_single} can be recast as 
	\begin{align}\label{thermal}
		\tilde{E}_W\left(A:A^c\right)=S\left(A\right)-S^{\text{Th}}\left(A\right)+\frac{1}{4G_N}\log(2)\, ,
	\end{align} 
	the first term represents the entanglement entropy, and the $S^{\text{Th}}$ term represents the thermal entropy. which shows that the HRT surface is getting wrapped by the horizon of the black hole. A similar kind of situation also arises when the boundary is getting pushed inside a finite cut-off according to the cut-off theory.

	\subsection{Finite Chemical Potential}

We now discuss the $\text{T}\bar{\text{T}}$ deformed BTZ black hole with a finite chemical potential $\Omega$ \cite{Banerjee:2024wtl}. In this case, the chiral functions $L$ and $\bar{L}$ are no longer equal, and in the FG coordinates the deformed metric is given as 
\begin{align}
		ds^2 = \frac{d\rho^2}{4 \rho^2} + \frac{1 + 4\mu^2L\bar{L} + L\bar{L}\rho(\rho + 8\mu + 4\mu^2L\bar{L}\rho)}{\rho(1 - 4\mu^2 L\bar{L})^2}dUdV \notag \\
		+ \frac{(\rho + 2\mu)(1 + 2\mu L\bar{L}\rho)}{\rho(1 - 4\mu^2 L\bar{L})^2}(LdU^2 + \bar{L}dV^2)\label{R_BTZ_metric_FG}
\end{align}
where $L$ and $\bar{L}$ are related to the outer and inner horizon radii $r_+$ and $r_-$ of the rotating BTZ black hole as
	\begin{align}\label{rprm}
		L = \frac{1}{4}(r_+ - r_-)^2\,,\quad \bar{L} = \frac{1}{4}(r_+ + r_-)^2\,.
	\end{align}
Through the coordinate transformations given as 
\begin{align}\label{r_BTZ_trans}
		\rho = \frac{r^2 - (L + \bar{L}) - \sqrt{(r^2 - L - \bar{L})^2 - 4L\bar{L}}}{2L\bar{L}}\,,\quad U = X + T\,,\quad V = X - T\,,
	\end{align}
    the deformed metric may be expressed as 
    \begin{align}\label{Rot_BTZ_Def_metric}
		ds^2 = \frac{r^2 \, dr^2}{(r^2 - r_+^2)(r^2 - r_-^2)} - \frac{4 (r^2 - r_+^2)(r_+ dT - r_- dX)^2}{r_h^2 (2 + r_h^2 \mu)^2} + \frac{4 (r^2 - r_-^2)(r_+ dX - r_- dT)^2}{r_h^2 (2 - r_h^2 \mu)^2}\, ,
	\end{align}
    where $r_h^2 = r_+^2 - r_-^2$. The above geometry may then be embedded in $\mathbb{R}^{(2,2)}$ using the embedding relations 
\begin{align}
		&X_{1} = \sqrt{\frac{r^{2} - r_{+}^{2}}{r_{+}^{2} - r_{-}^{2}}}\,
		\cosh\!\big[c(\mu)\big(r_{+}X - r_{-}T\big)\big],
		\qquad
		X_{2} = \sqrt{\frac{r^{2} - r_{-}^{2}}{r_{+}^{2} - r_{-}^{2}}}\,
		\sinh\!\big[d(\mu)\big(r_{+}T - r_{-}X\big)\big], \notag\\[4pt]
		&X_{3} = \sqrt{\frac{r^{2} - r_{+}^{2}}{r_{+}^{2} - r_{-}^{2}}}\,
		\sinh\!\big[c(\mu)\big(r_{+}X - r_{-}T\big)\big],
		\qquad
		X_{4} = \sqrt{\frac{r^{2} - r_{-}^{2}}{r_{+}^{2} - r_{-}^{2}}}\,
		\cosh\!\big[d(\mu)\big(r_{+}T - r_{-}X\big)\big],
		\label{Embedding-coordinates}
\end{align}
where 
\begin{align}
		c(\mu) = \frac{2}{2 + r_{h}^{2}\mu},
		\qquad\qquad
		d(\mu) = \frac{2}{2 - r_{h}^{2}\mu} \, .
\end{align}
We may further obtain the relations between the chiral functions and the inverse temperature and the chemical potential using the metric \cref{Rot_BTZ_Def_metric} as
\begin{align}\label{L_barL}
	L = \left( \frac{\beta_-(1-s)}{4\pi\mu} \right)^{2}, \quad 
	\bar{L} = \left( \frac{\beta_+(1-s)}{4\pi\mu} \right)^{2}.
	\end{align}
    where
\begin{align}
		\beta_\pm=\beta\left(1\pm\Omega\right), \qquad s = \sqrt{\,1 - \frac{8\pi^2 \mu}{\beta_+ \beta_-}}.
	\end{align}
This once again highlights the fact that in contrast to the conventional cutoff description, where the deformation in the bulk geometry manifests as a cutoff at some finite $r=r_c$, in the MBC formalism the deformation manifests as a modification to the inner and outer radii of the rotating BTZ black hole (as evident from \cref{rprm,L_barL}).

	\subsubsection{Two disjoint intervals}
	We consider two non-overlapping boosted intervals
$A=[(x_1,t_1),(x_2,t_2)]$ and $B=[(x_3,t_3),(x_4,t_4)]$
in a finite-temperature $\text{T}\bar{\text{T}}$-deformed CFT$_2$
with conserved charge $\Omega$.
The bulk dual is a rotating BTZ black hole, whose entanglement wedge
is bounded by geodesics anchored at the interval endpoints.
Using the embedding coordinates in \cref{Embedding-coordinates}
and the EWCS formula in \cref{EW-disj-formula},
We compute the entanglement wedge cross-section between $A$ and $B$,
with the deformed horizons identified via the undeformed inverse
temperature as in \cref{L_barL}.
 Utilizing these relations, we compute the EWCS in the corresponding deformed background. On the field theory side, the deformed result lacks a complete non-perturbative interpretation; therefore, we expand the bulk expression perturbatively around the deformation parameter. Up to first order in $\mu$, the result takes the following form:
	\begin{align}
		E_W(A:B) &= \frac{1}{4G_N} \cosh^{-1} \left[\frac{1 + \sqrt{\eta \bar{\eta}}}{\sqrt{(1 - \eta)(1 - \bar{\eta})}} \right] \notag \\
		&\quad + \frac{\mu \beta \pi^3 \sqrt{\eta \bar{\eta}}}{2G_N \beta_+^2 \beta_-^2 (\sqrt{\eta} + \sqrt{\bar{\eta}})}\left[ \bK_{21} + \bK_{43} - \bK_{32} - \bK_{41} \right] \notag\\
		&\quad + \frac{\mu \beta \pi^3}{2G_N \beta_+^2 \beta_-^2 (\sqrt{\eta} + \sqrt{\bar{\eta}})}\left[ \bK_{31} + \bK_{42} - \bK_{32} - \bK_{41} \right]\label{EW-disj}
	\end{align}
	where $\eta$ and $\bar{\eta}$ denote the cross ratios defined as
	\begin{align}
		\eta =\frac{\sinh \left(\frac{\pi  \left(x_{21}+t_{21}\right)}{\beta_+}\right) \sinh \left(\frac{\pi  \left(x_{43}+t_{43}\right)}{\beta_+}\right)}{\sinh \left(\frac{\pi  \left(x_{31}+t_{31}\right)}{\beta_+}\right) \sinh \left(\frac{\pi  \left(x_{42}+t_{42}\right)}{\beta_+}\right)}\,, \qquad \bar{\eta} =\frac{\sinh \left(\frac{\pi  \left(x_{21}-t_{21}\right)}{\beta_-}\right) \sinh \left(\frac{\pi  \left(x_{43}-t_{43}\right)}{\beta_-}\right)}{\sinh \left(\frac{\pi  \left(x_{31}-t_{31}\right)}{\beta_-}\right) \sinh \left(\frac{\pi  \left(x_{42}-t_{42}\right)}{\beta_-}\right)} \, .
	\end{align}
	
	Here, the first term represents the undeformed EWCS, while the second term captures the first-order correction induced by the $\text{T}\bar{\text{T}}$ deformation. For convenience, we introduce the following compact notation:
	\begin{align}
		\bK_{ij} = \cK_{ij} + \bar{\cK}_{ij} = (x_{ij} - \Omega t_{ij})\left[\coth\left(\frac{\pi(x_{ij} + t_{ij})}{\beta_+}\right) + \coth\left(\frac{\pi(x_{ij} - t_{ij})}{\beta_-}\right)\right]\label{bP-ij}\, .
	\end{align}
	
	The first-order correction obtained above precisely matches the field-theoretic computation in \cite{Basu:2024enr}, and it agrees with the cutoff-based holographic description discussed in \cite{Basu:2024enr} in the high-temperature regime.
	\subsubsection{Two adjacent intervals}
	We now consider two boosted intervals $A = [(x_1, t_1), (x_2, t_2)]$ and $B = [(x_2, t_2), (x_3, t_3)]$ in the boundary theory. This configuration corresponds to adjacent subsystems in the $\text{T}\bar{\text{T}}$-deformed CFT$_2$ at finite temperature with a conserved charge. To find the entanglement wedge cross-section in this scenario, we use the embedding coordinates given in \cref{Em_BTZ} and the same EWCS formula we used for non-zero conserved charge in \cref{EW-adj-formula}. Using this relation, we obtain the non-perturbative result in the deformed theory. Further, we replace the deformed horizons by the original, undeformed left and right moving inverse temperature given in \cref{L_barL}. Expanding the result up to first order around the deformation parameter, the corresponding entanglement wedge cross-section becomes as follows
	\begin{align}
		E_W(A:B) &= \frac{1}{8G_N} \log\left[\frac{2\beta_+}{\pi \epsilon_c} \frac{\sinh\left(\frac{\pi(x_{21}+t_{21})}{\beta_+}\right) \sinh\left(\frac{\pi(x_{32}+t_{32})}{\beta_+}\right)}{\sinh\left(\frac{\pi(x_{31}+t_{31})}{\beta_+}\right)}\right]\notag \\
		& + \frac{1}{8G_N} \log\left[\frac{2\beta_-}{\pi \epsilon_c} \frac{\sinh\left(\frac{\pi(x_{21}-t_{21})}{\beta_-}\right) \sinh\left(\frac{\pi(x_{32}-t_{32})}{\beta_-}\right)}{\sinh\left(\frac{\pi(x_{31}-t_{31})}{\beta_-}\right)}\right]\notag \\
		& + \frac{\pi^3 \beta \mu}{2G_N \beta_+^2 \beta_-^2}\left(\bK_{21} + \bK_{32} - \bK_{31}\right) - \frac{\pi^2 \mu}{2G_N \beta_+ \beta_-}\, .\label{EW-adj-corr}
	\end{align}
	Here, the first two logarithmic terms represent the undeformed EWCS for adjacent intervals in the rotating BTZ background, while the remaining terms correspond to the leading-order correction arising from the $\text{T}\bar{\text{T}}$ deformation. The quantity $\bK_{ij}$ is defined as in \cref{bP-ij}. If we drop the last term, then the first-order correction obtained here is consistent with the corresponding field-theoretic computation in \cite{Basu:2024enr}, and in the high-temperature limit $\beta<<|x_{ij}|$, it reproduces the cutoff-dependent result from the holographic description discussed therein.
	\subsubsection{A single interval}
	Finally, we consider the case of a single interval $A = [(0, 0), (\ell, t)]$. To compute the reflected entropy, we introduce auxiliary intervals $B_1 = [(-L, -T), (0, 0)]$ and $B_2 = [(\ell, t), (L, T)]$ such that $B = B_1 \cup B_2$ effectively approximates the complement $A^c$ in the limit $L \to \infty$. This construction parallels the zero angular momentum case, where the upper bound of the entanglement wedge cross section was determined using the relation \cref{UP_RE}. 
	In the present analysis, we employ the adjacent interval result derived earlier. Accordingly, we add the EWCS contributions corresponding to the adjacent intervals and then take the bipartite limit. Subsequently, by replacing the deformed horizon parameters with the left- and right-moving temperatures \cref{L_barL}, and expanding the resulting expression perturbatively around $\mu = 0$, we obtain the EWCS up to first order in the deformation parameter as follows
	\begin{align}
		E_W(A:B) &= \frac{1}{4G_N} \log \left(\frac{2\beta_+ \beta_- \sinh\left(\frac{\pi(\ell - t)}{\beta_-}\right) \sinh\left(\frac{\pi(\ell + t)}{\beta_+}\right)}{\pi^2 \epsilon_c^2} \right)\notag \\
		&+ \frac{\pi(\ell - t\Omega)}{2\beta G_N(\Omega^2 - 1)} + \frac{2\pi^3 \beta \mu (\ell - t\Omega)}{G_N \beta_+^2 \beta_-^2} \left[\coth\left(\frac{\pi(\ell - t)}{\beta_-}\right) + \coth\left(\frac{\pi(\ell + t)}{\beta_+}\right) - 2 \right] \notag \\
		&-\frac{\pi^2 \mu}{G_N \beta^2(1-\Omega^2)}\, .
	\end{align}
	Neglecting the last term, we observe that the remaining term proportional to $\mu$ agrees precisely with the field-theoretic result obtained in \cite{Basu:2024enr}. It also reproduces the cutoff-based description for the positive deformation parameter in the high-temperature regime. 
	
	It is important to note that the final expression exhibits a similar structure to \cref{thermal}, indicating that in the presence of a non-zero conserved charge, the horizon wraps the geodesic and contributes an additional thermal component to the entropy, given by
	\begin{align}
		S^{\text{Th}} = \frac{\pi(\ell - t\Omega)}{2\beta G_N(\Omega^2 - 1)}\, .
	\end{align}
	This wrapping effect highlights the significant role of the conserved charge, whose influence becomes increasingly prominent with time.

    
	\section{Entanglement Entropy in root-$\text{T}\bar{\text{T}}$ Deformed CFTs}\label{sec4}
    
	Having verified the validity of the MBC formalism for mixed state entanglement in $\text{T}\bar{\text{T}}$ deformed CFT$_2$s using the reflected entropy, we now investigate the entanglement entropy in root-$\text{T}\bar{\text{T}}$ deformed CFT$_2$s from both the field theoretic and bulk perspectives.
    For a small deformation parameter $\mu$, the theory admits a perturbative description around the undeformed CFT
    seed action\cite{Ferko:2022cix,Ebert:2023tih}
	\begin{align}
		\mathcal{I} = \mathcal{I}_{\text{CFT}} + \mu \int_{\mathcal{M}} \sqrt{\text{T}\bar{\text{T}} - \Theta^2} \, ,
	\end{align}
    where $T$, $\bar{T}$ and $\Theta$ defines the stress tensor components. In this scenario, we may now obtain physical quantities using perturbation techniques around the seed theory. Here, we particularly focus on the entanglement entropy for a single interval in a thermal CFT$_2$ both without and with a conserved charge, and subsequently study the holographic description using mixed Dirichlet–Neumann boundary conditions.

    \subsection{Entanglement entropy}
    Here, we discuss the computation of entanglement entropy for a single interval $A$ defined on a thermal cylinder, whereas for a finite conserved charge, the cylinder is a twisted cylinder. Now, the R\'enyi entropy associated with a subsystem  on the manifold $\mathcal{M}$ is defined as
	\begin{align}
		S_n(A) = \frac{1}{1 - n} \log \frac{Z_n(A)}{Z^n} \, , \qquad 
		S(A) = \lim_{n \to 1} S_n(A) \, ,
	\end{align}
	where $Z_n(A)$ denotes the partition function evaluated on the $n$-sheeted Riemann surface $\mathcal{M}_n$, which is constructed by cyclically gluing $n$ copies of the original manifold $\mathcal{M}$ along the subsystem $A$ according to the replica prescription \cite{Calabrese:2009qy}. The quantity $Z$ represents the partition function on a single copy of $\mathcal{M}$. The entanglement entropy is then obtained by analytically continuing the R\'enyi index $n$ to real values and taking the limit $n \to 1$. This procedure effectively captures the quantum correlations between $A$ and its complement, providing a key probe of the underlying structure of the deformed theory.
	
	To compute the entanglement entropy, we first evaluate the ratio $Z_n / Z^n$. Since the deformation parameter $\mu$ is small, we can expand the expression perturbatively in powers of $\mu$, yielding
	\begin{align}\label{partition_ratio}
		\frac{Z_n(A)}{Z^n} = 
		\frac{\int_{\mathcal{M}^n} e^{-\mathcal{I}_{\text{CFT}}} 
			\left( 1 - \mu \int_{\mathcal{M}^n} \sqrt{\text{T}\bar{\text{T}} - \Theta^2} + \mathcal{O}(\mu^2) \right)}
		{\left[ \int_{\mathcal{M}} e^{-\mathcal{I}_{\text{CFT}}} 
			\left( 1 - \mu \int_{\mathcal{M}} \sqrt{\text{T}\bar{\text{T}} - \Theta^2} + \mathcal{O}(\mu^2) \right) \right]^n} \, .
	\end{align}
	It is important to note that, when a CFT is defined on a flat manifold, any correlation function involving the trace of the energy-momentum tensor vanishes, i.e., $\langle T^\mu_{\ \mu} \ldots \rangle = 0$. In what follows, we consider the manifold $\mathcal{M}$ to be a cylinder. In this setup, $\sigma$ denotes the twist operator that identifies the fields when two adjacent copies of $\mathcal{M}$ are joined along the interval $A$
	\footnote{In \cref{partition_ratio}, the square root term is expanded using a binomial expansion, leading to contributions from the $\text{T}\bar{\text{T}}$ operator, which we collectively denote by $X$.}.
	Accordingly, we have
	\begin{align}
		\int_{\mathcal{M}} e^{-\mathcal{I}_{\text{CFT}}} \, \Theta^2 X &\sim \langle \Theta^2 X \rangle_{\mathcal{M}} = 0 \, ,\\
		\int_{\mathcal{M}_n} e^{-\mathcal{I}_{\text{CFT}}} \, \Theta^2 X &\sim \langle \Theta^2 X \sigma \rangle_{\mathcal{M}_n} = 0 \, .
	\end{align}
	Furthermore, using the relation
	\begin{align}
		\int_{\mathcal{M}_n} \langle \sqrt{\text{T}\bar{\text{T}}} \rangle_{\mathcal{M}_n}
		= n \int_{\mathcal{M}} \langle \sqrt{\text{T}\bar{\text{T}}} \rangle \, ,
	\end{align}
	we obtain
	\begin{align}
		\frac{Z_n(A)}{Z^n} 
		= \frac{\int_{\mathcal{M}^n} e^{-\mathcal{I}_{\text{CFT}}}}{\left[\int_{\mathcal{M}} e^{-\mathcal{I}_{\text{CFT}}}\right]^n}
		\left( 1 - \mu \int_{\mathcal{M}^n} \langle \sqrt{\text{T}\bar{\text{T}}} \rangle_{\mathcal{M}_n} 
		+ n \mu \langle \sqrt{\text{T}\bar{\text{T}}} \rangle 
		+ \mathcal{O}(\mu^n) \right) \, .
	\end{align}
	Substituting this result into the R\'enyi entropy formula, we obtain the first-order correction to the entanglement entropy for the subsystem $A$ on the manifold $\mathcal{M}$ due to the root-$\text{T}\bar{\text{T}}$ deformation,
	\begin{align}\label{delta_S1}
		\delta S_n(A) = 
		\frac{-n\mu}{1 - n} 
		\int_{\mathcal{M}} 
		\left[ 
		\langle \sqrt{\text{T}\bar{\text{T}}} \rangle_{\mathcal{M}_n} 
		- \langle \sqrt{\text{T}\bar{\text{T}}} \rangle_{\mathcal{M}} 
		\right] \, .
	\end{align}
	Taking the limit $n \to 1$, one can then extract the corresponding correction to the entanglement entropy. A similar analysis for the standard $\text{T}\bar{\text{T}}$ deformation was presented in \cite{Chen:2018eqk}.
	\subsection{Finite Temperature}	
	In this subsection, we consider a two-dimensional CFT deformed by the root-$\text{T}\bar{\text{T}}$ operator at a finite temperature, characterized by the inverse temperature $\beta$. The theory is defined on a cylindrical manifold $\mathcal{M}$, where the Euclidean time direction is compactified with circumference $\beta$, while the spatial direction remains non-compact. The complex coordinates on the cylinder are denoted by
	\begin{align}
		w = x + i \tau \, , \qquad \bar{w} = x - i \tau \, .
	\end{align}
	On this background, we consider a boosted interval $A = \left[\left(x_2, t_2\right), \left(x_1, t_1\right)\right]$ along the spatial direction.
	
	\vspace{0.2cm}
	
	To proceed, we perform the conformal mapping
	\begin{align}\label{conformal_map1}
		z = e^{\frac{2 \pi}{\beta} w} \, ,
	\end{align}
	which maps the cylindrical manifold $\mathcal{M}$ to the complex plane $\mathcal{C}$. Under this map, the holomorphic component of the stress tensor transforms according to the standard conformal transformation law,
	\begin{align}\label{T_trans}
		T(w) = \left( \frac{d z}{d w} \right)^2 T(z) + \frac{c}{12} \{ z , w \} \, ,
	\end{align}
	where $\{z, w\}$ is the Schwarzian derivative defined as
	\begin{align}\label{Swarz_deri}
		\{ z, w \} = \frac{z^{\prime\prime\prime}}{z^{\prime}} - \frac{3}{2} \left( \frac{z^{\prime\prime}}{z^{\prime}} \right)^2 \, .
	\end{align}
	A similar transformation holds for the antiholomorphic component $\bar{T}(\bar{w})$.  
	
	\vspace{0.2cm}
	
	Using these relations, and applying a binomial expansion to the square root in the deformation term\footnote{We employ the binomial expansion for the square root term and impose the condition $\langle T(z) \rangle_{\mathcal{C}} = 0$ on the plane.}, we obtain the following expectation value of the composite operator on the cylinder:
	\begin{align}\label{TTbar_M}
		\left\langle \sqrt{\text{T}\bar{\text{T}}(w, \bar{w})} \right\rangle_{\mathcal{M}} 
		= \frac{c}{12} \sqrt{ \{ z, w \} \{ \bar{z}, \bar{w} \} }
		= \frac{c}{12} \frac{2 \pi^2}{\beta^2} \, .
	\end{align}
	
	\vspace{0.2cm}
	
	To evaluate $\left\langle \sqrt{\text{T}\bar{\text{T}}(w, \bar{w})} \right\rangle_{\mathcal{M}_n}$ on the $n$-sheeted Riemann surface $\mathcal{M}_n$, we must introduce an additional conformal map. The transformation in \cref{conformal_map1} maps the interval $A$ on the cylinder to an interval $A'$ on the plane $\mathcal{C}$, such that the $n$-sheeted surface $\mathcal{M}_n$ corresponds to a branched cover $\mathcal{C}_n$, constructed by gluing $n$ copies of the complex plane along $A'$. The surface $\mathcal{C}_n$ can then be mapped back to the complex plane $\mathcal{C}$ via
	\begin{align}
		w' \to z = \left( \frac{ w' - e^{\frac{2\pi (x_1 + i\tau_1)}{\beta}} }{ w' - e^{\frac{2\pi (x_2 + i\tau_2)}{\beta}} } \right)^{\frac{1}{n}} \, ,
	\end{align}
	where a detailed discussion of this mapping can be found in \cite{Calabrese:2009qy}. Combining the two conformal maps and applying the transformation law \cref{T_trans}, we obtain the corresponding expectation value on the replicated manifold:
	\begin{align}\label{TTbar_Mn}
		\left\langle \sqrt{\text{T}\bar{\text{T}}(w, \bar{w})} \right\rangle_{\mathcal{M}_n}
		= \frac{c}{12} \sqrt{ \{ z, w \} \{ \bar{z}, \bar{w} \} } \, .
	\end{align}
	
	\vspace{0.2cm}
	
	Since we are ultimately interested in the entanglement entropy, we take the $n \to 1$ limit. To this end, we expand the Schwarzian derivatives around $n = 1$:
	\begin{align}
		\sqrt{ \{ z, w \} \{ \bar{z}, \bar{w} \} } =\frac{2\pi^2}{\beta^2}+\frac{2\pi^2}{\beta^2}\left(1-n\right)\left(\mathcal{F}+\bar{\mathcal{F}}\right)+ \mathcal{O}\left( (n - 1)^2 \right)\, ,
	\end{align}
	where the functions $\mathcal{F}$ and $\bar{\mathcal{F}}$ are given by
	\begin{align}
		\mathcal{F} &=
		\frac{\left(e^{\frac{\pi(i(\tau_1 - \tau_2) + x_1 - x_2)}{\beta}} 
			- e^{-\frac{\pi(i(\tau_1 - \tau_2) + x_1 - x_2)}{\beta}}\right)^2}
		{\left(e^{\frac{\pi(-i\tau_1 + w - x_1)}{\beta}} 
			- e^{-\frac{\pi(-i\tau_1 + w - x_1)}{\beta}}\right)^2
			\left(e^{\frac{\pi(-i\tau_2 + w - x_2)}{\beta}} 
			- e^{-\frac{\pi(-i\tau_2 + w - x_2)}{\beta}}\right)^2} \, , \notag\\
		\bar{\mathcal{F}} &=
		\frac{\left(e^{\frac{\pi(i(\tau_1 - \tau_2) + x_1 - x_2)}{\beta}} 
			- e^{-\frac{\pi(i(\tau_1 - \tau_2) + x_1 - x_2)}{\beta}}\right)^2}
		{\left(e^{\frac{\pi(-i\tau_1 + \bar{w} - x_1)}{\beta}} 
			- e^{-\frac{\pi(-i\tau_1 + \bar{w} - x_1)}{\beta}}\right)^2
			\left(e^{\frac{\pi(-i\tau_2 + \bar{w} - x_2)}{\beta}} 
			- e^{-\frac{\pi(-i\tau_2 + \bar{w} - x_2)}{\beta}}\right)^2} \, .
	\end{align}
	Substituting these expressions into \cref{TTbar_Mn}, we obtain the expectation value on $\mathcal{M}_n$:
	\begin{align}
		\left\langle \sqrt{\text{T}\bar{\text{T}}(w,\bar{w})} \right\rangle_{\mathcal{M}_n}
		=  \frac{c}{6} \frac{\pi^2}{\beta^2}
		\left[ 1 + (1-n)(\mathcal{F} + \bar{\mathcal{F}}) \right] \, .
	\end{align}
	
	Finally, substituting the results from \cref{delta_S1,TTbar_M,TTbar_Mn} and taking the limit $n \to 1$, we find the correction to the entanglement entropy for a single interval $A$ due to the root-$\text{T}\bar{\text{T}}$ deformation at finite temperature as follows
	\begin{align}
		\delta S(A)
		= - \mu \, \frac{c}{6} \, \frac{\pi^2}{\beta^2} 
		\int_{\mathcal{M}} \left( \mathcal{F} + \bar{\mathcal{F}} \right) \, .
	\end{align}
	A similar type of integral was computed in \cite{Chen:2018eqk} for the standard $\text{T}\bar{\text{T}}$-deformed CFT, allowing one to determine the corresponding correction to the entanglement entropy for a single interval in the deformed thermal state. The result takes the following form:
	\begin{align}
		\delta S(A) = -\frac{\mu \pi^2 c \, x_{21}}{6 \beta}
		\left[
		\coth\!\left( \frac{\pi}{\beta}(x_{21} + t_{21}) \right)
		+ 
		\coth\!\left( \frac{\pi}{\beta}(x_{21} - t_{21}) \right)
		\right] \, .
	\end{align}
	In obtaining the above expression, an analytic continuation from the Euclidean to Lorentzian signature was performed, where the Euclidean time coordinate is Wick rotated as $\tau \to i t$. This continuation ensures that the resulting expression correctly describes the entanglement entropy dynamics in the Lorentzian framework of the deformed CFT.
	\subsection{Finite Temperature with Conserved Charge}
	
	In the previous section, we discussed the case of a thermal CFT defined on a cylinder without any conserved charge. We now extend our analysis to a thermal CFT that possesses a conserved charge. The presence of this charge effectively twists the geometry, such that the subsystem $A = \left[\left(x_2, t_2\right), \left(x_1, t_1\right)\right]$ now lies on a \emph{twisted cylinder}. In this setup, the compactification direction makes a finite angle with the Euclidean time direction, and the conserved charge is directly related to this twist parameter.\footnote{The structure and detailed geometry of a twisted cylinder are discussed in \cite{Basu:2024enr}.}
	
	To map this twisted cylinder to the complex plane $\mathcal{C}$, we employ the following conformal transformation:
	\begin{align}\label{twsited_map}
		z \to e^{\frac{2 \pi}{\beta_+} w}, \qquad 
		\bar{z} \to e^{\frac{2 \pi}{\beta_-} \bar{w}} \, ,
	\end{align}
	where the parameters $\beta_\pm = \beta (1 \pm i \Omega_E)$ encode both the inverse temperature $\beta$ and the conserved charge $\Omega_E$. The total circumference of the twisted cylinder is thus modified to $\beta \sqrt{1 + \Omega_E^2}$. Using the transformations \cref{T_trans,Swarz_deri,twsited_map}, the expectation value of the root-$\text{T}\bar{\text{T}}$ operator on this manifold $\mathcal{M}$ becomes
	\begin{align}
		\left\langle \sqrt{\text{T}\bar{\text{T}}(w,\bar{w})} \right\rangle_{\mathcal{M}} 
		= \frac{c}{12} \sqrt{ \{z,w\} \, \{\bar{z},\bar{w}\} } 
		= \frac{c}{12} \frac{2 \pi^2}{\beta_+ \beta_-} \, .
	\end{align}
	To compute the expectation value of the root-$\text{T}\bar{\text{T}}$ operator on the $n$-replica manifold $\mathcal{M}_n$, we follow a procedure analogous to the zero-charge case. Since the interval now resides on a twisted cylinder, the conformal map in \cref{twsited_map} sends the manifold $\mathcal{M}_n$ to a replicated surface $\mathcal{C}_n$, which consists of $n$ copies of the complex plane $\mathcal{C}$ connected along the interval. This is achieved via the following mapping
	\begin{align}\label{twisted_map2}
		w = \left( \frac{w' - e^{\frac{2\pi}{\beta_+}(x_1 + i\tau_1)}}{w' - e^{\frac{2\pi}{\beta_+}(x_2 + i\tau_2)}} \right)^{\!\frac{1}{n}}, 
		\qquad 
		\bar{w} = \left( \frac{\bar{w}' - e^{\frac{2\pi}{\beta_-}(x_1 - i\tau_1)}}{\bar{w}' - e^{\frac{2\pi}{\beta_-}(x_2 - i\tau_2)}} \right)^{\!\frac{1}{n}} \, ,
	\end{align}
	which maps $\mathcal{C}_n$ back to the single-sheeted plane $\mathcal{C}$.
	
	Next, using \cref{twisted_map2}, one can evaluate the Schwarzian derivatives $\sqrt{\{z,w\}}$ and $\sqrt{\{\bar{z},\bar{w}\}}$. Expanding around $n = 1$, we obtain
	\begin{align}
	\sqrt{ \{ z, w \} \{ \bar{z}, \bar{w} \} } =\frac{2\pi^2}{\beta_+ \beta_-}+\frac{2\pi^2}{\beta_+ \beta_-}\left(1-n\right)\left(\mathcal{G}+\bar{\mathcal{G}}\right)+ \mathcal{O}\left( (n - 1)^2 \right)
	\end{align}
	where the functions $\mathcal{G}$ and $\bar{\mathcal{G}}$ are given by
	\begin{align}
		\mathcal{G} &= 
		\frac{ \left( e^{\frac{\pi (i(\tau_1 - \tau_2) + x_1 - x_2)}{\beta_+}} - e^{-\frac{\pi (i(\tau_1 - \tau_2) + x_1 - x_2)}{\beta_+}} \right)^2 }
		{ \left( e^{\frac{\pi (-i\tau_1 + w - x_1)}{\beta_+}} - e^{-\frac{\pi (-i\tau_1 + w - x_1)}{\beta_+}} \right)^2 
			\left( e^{\frac{\pi (-i\tau_2 + w - x_2)}{\beta_+}} - e^{-\frac{\pi (-i\tau_2 + w - x_2)}{\beta_+}} \right)^2 } \, , \notag\\
		\bar{\mathcal{G}} &= 
		\frac{ \left( e^{\frac{\pi (i(\tau_1 - \tau_2) + x_1 - x_2)}{\beta_-}} - e^{-\frac{\pi (i(\tau_1 - \tau_2) + x_1 - x_2)}{\beta_-}} \right)^2 }
		{ \left( e^{\frac{\pi (-i\tau_1 + \bar{w} - x_1)}{\beta_-}} - e^{-\frac{\pi (-i\tau_1 + \bar{w} - x_1)}{\beta_-}} \right)^2 
			\left( e^{\frac{\pi (-i\tau_2 + \bar{w} - x_2)}{\beta_-}} - e^{-\frac{\pi (-i\tau_2 + \bar{w} - x_2)}{\beta_-}} \right)^2 } \, .
	\end{align}
	
	Using \cref{delta_S1}, we can then express the first-order correction to the entanglement entropy in the presence of a conserved charge $\Omega$ as
	\begin{align}
		\delta S(A) = -\mu \frac{c}{6} \frac{\pi^2}{\beta_+ \beta_-} 
		\int_{\mathcal{M}} \left( \mathcal{G} + \bar{\mathcal{G}} \right) \, .
	\end{align}
	Here, $\Omega = -i\Omega_E$ is the Lorentzian continuation of the conserved charge. The computation of the above integral requires an appropriate coordinate transformation, after which it can be evaluated explicitly. Subsequently, one can revert to the original coordinate system using the inverse map.\footnote{The detailed computation of this integral can be found in \cite{Basu:2024enr}.} 
	
	Finally, taking the Euclidean time to the Lorentzian by Wick rotation $t=-i\tau$, the correction to the entanglement entropy for a single interval in the presence of a conserved charge takes the compact form
	\begin{align}\label{CFT_rTTbar1}
		\delta S(A) = -\frac{\mu \pi^2 c \, \beta (x_{21} - \Omega t_{21})}{6 \beta_+ \beta_-}
		\left[
		\coth\!\left( \frac{\pi}{\beta_+}(x_{21} + t_{21}) \right)
		+ 
		\coth\!\left( \frac{\pi}{\beta_-}(x_{21} - t_{21}) \right)
		\right] \, .
	\end{align}
	This result encapsulates the nontrivial interplay between temperature, conserved charge, and the root-$\text{T}\bar{\text{T}}$ deformation, revealing how such deformations modify the entanglement structure of the thermal state in a charged background.
	\section{Holographic Entanglement Entropy in root-$\text{T}\bar{\text{T}}$ Deformed Theories}\label{sec5}
	In this part, we will discuss the correction to the entanglement entropy for a single interval in the bulk deformed BTZ background for two cases. In the first case, we consider the scenario where no chemical potential or angular momentum is present; in the second example, we will consider the presence of angular momentum. 
	\subsection{Vanishing Chemical Potential}
	We begin with the case of vanishing chemical potential, where the dual bulk geometry corresponds to a non-rotating BTZ black hole corresponding $\Omega=0$. In this case, we allow the chiral functions $L$ and $\bar{L}$ set to equal. Now, using the deformation relation derived in \cref{root_UV} and applying it to the Fefferman--Graham metric, we obtain the following deformed geometry
	\begin{align}
		ds^2 &= \frac{d\rho^2}{4 \rho^2} - \frac{\sinh(\mu) + L \rho (L \rho \sinh(\mu) - 2 \cosh(\mu))}{2 \rho} \left[(dU)^2 + (dV)^2\right] \\
		&\quad + \frac{ \left[\cosh(\mu) + L^2 \rho^2 \cosh(\mu) - 2L \rho \sinh(\mu)\right]}{\rho}dU dV\, ,
	\end{align}
	After performing a coordinate transformation \cref{BTZ trans}, the metric becomes as follows 
	\begin{align}\label{BTZ_def_rTTbar}
		ds^2 = -\frac{dr^2}{4L - r^2} + e^{\mu}(4L - r^2) dT^2 + e^{-\mu} r^2 dX^2\, ,
	\end{align}
     here we can observe for $\mu\to 0$, this metric becomes standard BTZ black hole with $r_h=2\sqrt{L}$. And this metric can be recast using the embedding coordinates \cref{Em_BTZ}, but for root$-\text{T}\bar{\text{T}}$ deformation the $\mu$ dependent functions are
	\begin{align}
		a\left(\mu\right)=e^{-\frac{\mu}{2}}\quad\quad b\left(\mu\right)=e^{\frac{\mu}{2}}\, .
	\end{align} 
	Now, in this background we consider a single boundary interval $A = [(x_1, t_1), (x_2, t_2)]$ at the boundary, which corresponds to a boosted subsystem in the thermal CFT. The dual entanglement entropy is obtained using the Ryu-Takayanagi prescription, where one computes the length of the bulk geodesic anchored at the endpoints of the interval in the deformed geometry.

	Applying this embedding coordinates and subsequently using the Ryu--Takayanagi formula, we evaluate the entanglement entropy in the deformed geometry. In this setup, the Euclidean periodicity of the background is also modified by the deformation. Consequently, the inverse temperature $\beta$ of the undeformed theory becomes related to the bulk parameter $L$ and the deformation parameter $\mu$ as
	\begin{align}\label{root_ttbar_temp}
		L = \frac{\pi^2 e^{-\mu}}{\beta^2} \, .
	\end{align}
	
	Using these relations, the geodesic length is first expressed in terms of $L$, and then $L$ is eliminated in favor of $\beta$, leading to a $\mu$-dependent expression. Expanding perturbatively in small $\mu$, the first-order correction to the entanglement entropy is found to be
	\begin{align}
		\delta S(A) 
		= \mu \frac{\pi c}{6\beta} \, x_{21} 
		\left[
		\coth \!\left( \frac{\pi (x_{21} + t_{21})}{\beta} \right)
		+ 
		\coth \!\left( \frac{\pi (x_{21} - t_{21})}{\beta} \right)
		\right]
		+ \frac{\mu}{4G} \, .
	\end{align}
	If we neglect the constant term for high temperature limit, and replace $\mu\to-\pi\mu$, then the first contribution precisely matches the field-theoretic result derived earlier \cref{CFT_rTTbar1}, after applying the Brown–Henneaux relation \cite{Brown:1986nw}. This agreement provides a strong consistency check of the bulk–boundary correspondence in the root-$\text{T}\bar{\text{T}}$ deformed background in the absence of a conserved charge.
	\subsection{Finite Chemical Potential}
	Next, we turn to the case of finite chemical potential or non-zero angular momentum, where the bulk geometry corresponds to a rotating BTZ black hole, and the boundary theory exhibits distinct left and right-moving temperatures. In FG coordinates, such a configuration the chiral functions may arise under the condition $L \neq \bar{L}$, and the corresponding $\text{root-}\text{T}\bar{\text{T}}$ deformed geometry takes the form  
	\begin{align}
		ds^2 &= \frac{d\rho^2}{4 \rho^2} - \frac{\sinh(\mu) \sqrt{\frac{L}{\bar{L}}} (L \bar{L} \rho^2 + 1) - 2L \rho \cosh(\mu)}{2 \rho} \left[(dU)^2 + (dV)^2\right] \nonumber \\
		&\quad + \frac{dU\, dV \left[\cosh(\mu)(L \bar{L} \rho^2 + 1) - \rho \sinh(\mu)(\sqrt{L} + \sqrt{\bar{L}})\right]}{\rho}\, .
	\end{align}
	
	Further, applying the coordinate transformation \cref{r_BTZ_trans} to the above expression, we obtain the deformed metric as  
	\begin{align}\label{rott_ttbar_metric_rotating}
		ds^2 = \frac{r^2 dr^2}{(r^2 - r_-^2)(r^2 - r_+^2)} - e^{\mu} (r^2 - r_+^2) \frac{(r_+ dT - r_- dX)^2}{r_+^2 - r_-^2} + e^{-\mu} (r^2 - r_-^2) \frac{(r_- dT - r_+ dX)^2}{r_+^2 - r_-^2}\, ,
	\end{align} 
    here $L$ and $\bar{L}$ is associated with similar fashion as it was for $\text{T}\bar{\text{T}}$ deformation. Here we can notice for $\mu\to0$ the metric represents a rotating BTZ black hole, and the undeformed thermodynamical parameters can be expressed for this deformed geometry as follows 
    \begin{align}\label{Beta_Omega_root}
		\beta = \frac{\pi}{2} \left(\frac{1}{\sqrt{L}} + \frac{1}{\sqrt{\bar{L}}}\right)e^{-\mu/2}, 
		\qquad 
		\Omega = \frac{\sqrt{\bar{L}} - \sqrt{L}}{\sqrt{\bar{L}} + \sqrt{L}}\, ,
	\end{align}
    where the deformation modifies the Euclidean periodicity, leading to a change in the temperature of the deformed rotating BTZ black hole. Alternatively \cref{Beta_Omega_root} can be expressed in terms of the left- and right-moving inverse temperatures using  
	\begin{align}\label{L_barL_betap_beta_-root_TTbar}
		L = \frac{\pi^2 e^{-\mu}}{\beta_+^2}, 
		\qquad 
		\bar{L} = \frac{\pi^2 e^{-\mu}}{\beta_-^2}\, ,
	\end{align}
	which clearly shows how the deformation affects the thermodynamic parameters through $\mu$ and the chiral sectors of the boundary theory.  
    
	Now we can obtain the deformed metric using the embedding coordinates used earlier \cref{Embedding-coordinates}, in this case, the $\mu$ dependent functions are 
		\begin{align}
		c\left(\mu\right)=e^{-\frac{\mu}{2}}\quad\quad d\left(\mu\right)=e^{\frac{\mu}{2}}\, .
	\end{align} 
	
	Now, we consider a boosted interval $A = [(x_1, t_1), (x_2, t_2)]$ on the boundary. To compute the entanglement entropy, we employ the Ryu–Takayanagi formula and $\mu$ dependent embedding coordinates
	. Finally by replacing the bulk parameters $L$ and $\bar{L}$ using \cref{L_barL_betap_beta_-root_TTbar}, we obtain the following expression that represents the entanglement entropy for a boosted single interval
	\begin{align}\label{HEE_rootTTbar}
	S&=\frac{1}{4G}\cosh^{-1}\!\left[ 
	-\frac{
		A
		\cosh\!\left( \dfrac{2\pi e^{-\mu}\big(x_{21}-\Omega\,t_{21} \big)}{\beta(1-\Omega^2 )} \right)
		+ B
		\cosh\!\left( \dfrac{2\pi\big(t_{21} - \Omega\,x_{21}\big)}{\beta(1-\Omega^2 )} \right)
	}{
		4\pi^2 u_c^2 (\Omega^2 - 1)
	}
	\right] \, ,
		\end{align}
		here $u_c$ is the UV cut off, and we define $A$ and $B$ as follows
		\begin{align}
			A=	\big(\beta^2 e^{\mu}(\Omega^2 - 1)^2 - 4\pi^2 u_c^2 \Omega^2\big)\,,\quad  B=\big(4\pi^2 u_c^2 - \beta^2 e^{\mu}(\Omega^2 - 1)^2\big)\notag\, .
		\end{align}
	This expression represents the total HEE in the root-$\text{T}\bar{\text{T}}$ deformed geometry. Since there is no non-perturbative correction from the field theoretic perspective, we perform a perturbative expansion around $\mu$, where the first-order contribution takes the form after using the Brown Honex relation
	\begin{align}
		\delta S(A) = \mu \frac{\pi c}{6(\beta_+ \beta_-)^{1/2}} \frac{(x_{21} - \Omega t_{21})}{\sqrt{1 - \Omega^2}} \left[\coth \left(\frac{\pi(x_{21} + t_{21})}{\beta_+}\right) + \coth \left(\frac{\pi(x_{21} - t_{21})}{\beta_-}\right)\right] + \frac{\mu}{4G}\, .
	\end{align}
	Neglecting the constant term for high temperature limit, we find that this expression matches precisely with the field-theoretic result obtained earlier in \cref{CFT_rTTbar1}.
	
	If we analyze the total non-perturbative result \cref{HEE_rootTTbar}, several important observations can be made. To ensure a real value of the total HEE, the term inside the $\cosh^{-1}$ must be greater than one. In this case, if we consider a purely spacelike interval, the deformation parameter must satisfy the bound
	\begin{align}
		\mu \le \log\left(\frac{1}{\Omega}\right) \notag\, ,
	\end{align}
	which indicates that for a non-rotating BTZ black hole, the deformation parameter has no restriction. However the upper bound of the deformation parameter decreases with increasing angular momentum.
	
	Similarly in the deformed geometry we can observe that the non-perturbative timelike HEE can be real, and for this to occur, the deformation parameter must satisfy
	\begin{align}
		\mu \le \log\left(\Omega\right) \notag\, .
	\end{align}
	This expression also implies that, to obtain a real timelike HEE in the deformed geometry, the deformation parameter must be negative. It further suggests that for a non-rotating BTZ black hole, the deformation parameter moves into an undefined region, indicating that such a situation arises only in the presence of angular momentum in the theory.
	
	Here we plot the total HEE as a function of the deformation parameter for different subsystem sizes. In the left panel, we show the result for the spacelike interval. In contrast, in the right panel, we present the variation of the real part of HEE with the deformation parameter for the timelike interval. 
	\begin{figure}[ht]
		\begin{center}
			\includegraphics[width=0.49\textwidth]{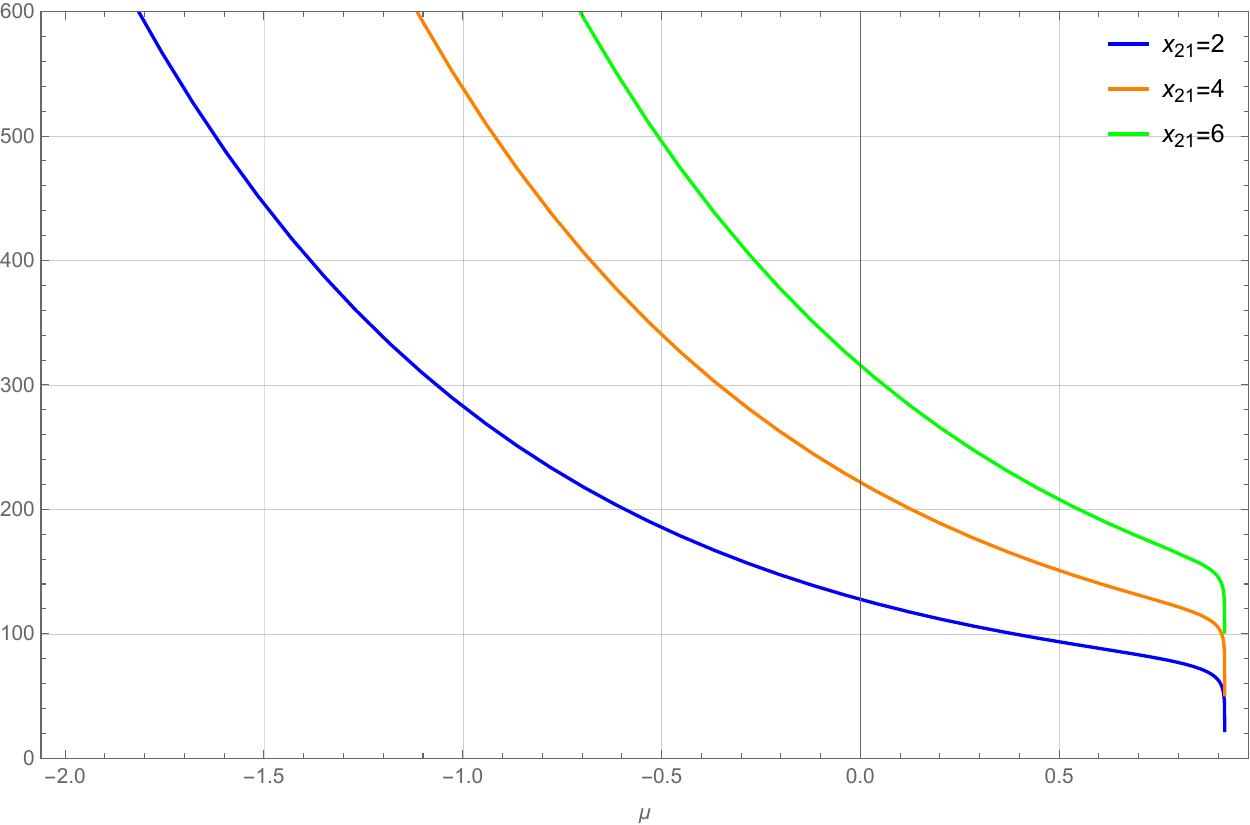}
			\includegraphics[width=0.49\textwidth]{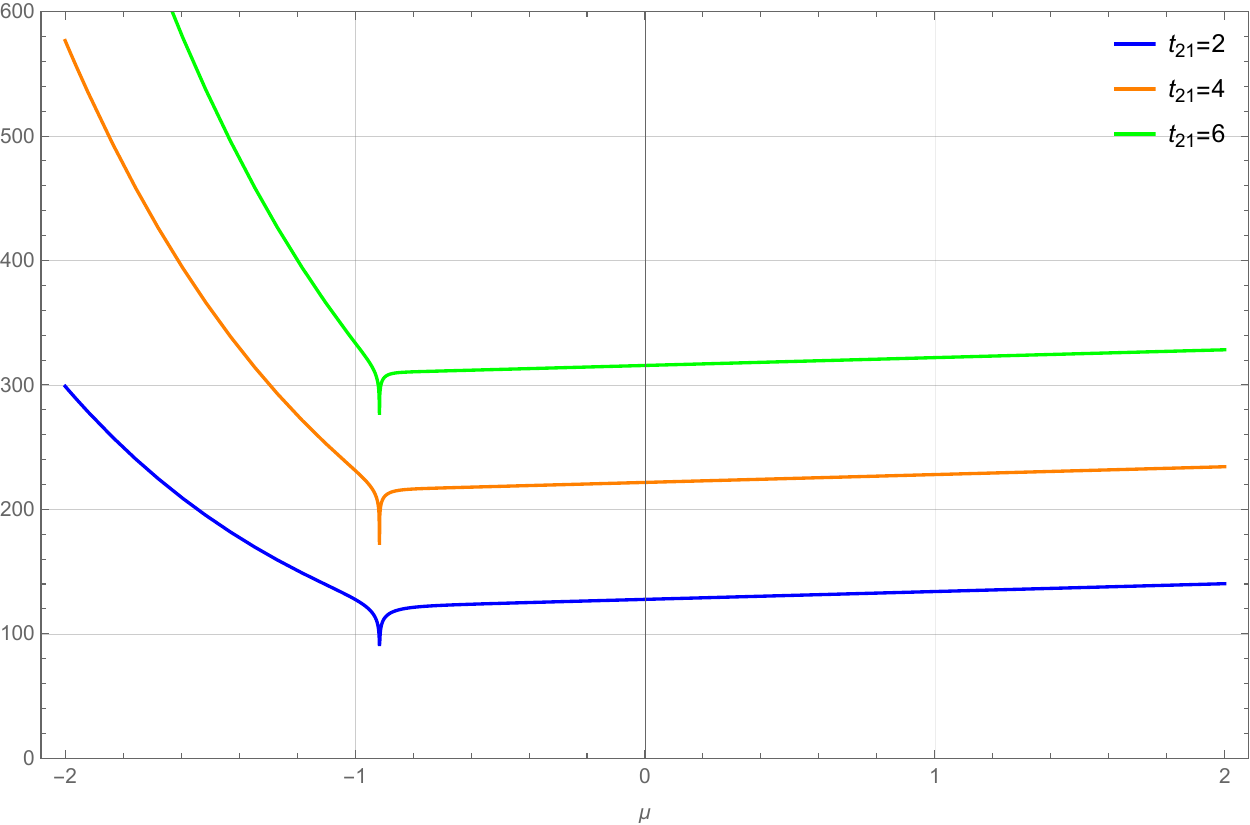}
			\caption{$\beta=1,~\Omega=0.4,~c=12\pi,~u_c=0.01$}
			\label{QvsMu}
		\end{center}
	\end{figure}
	
	In the left panel, we observe that at a certain value of $\mu$, the total HEE becomes zero and is independent of the subsystem size. In the right panel, the real part of the HEE also vanishes at a specific value of $\mu=\log(\Omega)$; the timelike HEE becomes complex for larger $\mu$ and remains real for smaller values.
	\section{QNEC in deformed theory}\label{sec6}
	In the following section, we study the Quantum Null Energy Condition (QNEC) in the context of $\text{T}\bar{\text{T}}$ and rooot-$\text{T}\bar{\text{T}}$ deformed quantum field theories. The QNEC establishes a fundamental relationship between the stress-energy tensor and variations of entanglement entropy. In a general two-dimensional quantum field theory, the QNEC takes the form
	\begin{align}\label{QNEC_QFT}
		\mathcal{Q}_{\pm} \equiv 2\pi \langle T_{\pm\pm} \rangle - S^{\prime\prime}_{\text{ent}} \geq 0,
	\end{align}
	where $T_{\pm\pm}$ are the null components of the stress tensor in the deformed geometry. For conformal field theories, the relation is
	\begin{align}\label{QNEC_CFT}
		\mathcal{Q}_{\pm} \equiv 2\pi \langle T_{\pm\pm} \rangle - S^{\prime\prime}_{\text{ent}} - \frac{6}{c} S^{\prime 2}_{\text{ent}} \geq 0,
	\end{align}
	with $S^{\prime}_{\text{ent}} = \partial_{\pm}S_{\text{ent}}$ representing the variation of entanglement entropy along null directions, and $c$ being the central charge.
	While modified versions of QNEC have been proposed for small deformations \cite{Banerjee:2024wtl}, the non-perturbative regime remains less understood. In this section, we take the general QFT version (\ref{QNEC_QFT}) as our starting point and investigate the QNEC under finite deformations. 
	\subsection{QNEC for $\text{T}\bar{\text{T}}$}
	
	In \cite{Banerjee:2024wtl}, the authors computed the HEE and obtained the QNEC for spacelike intervals. Here, we aim to generalize this analysis for boosted intervals with both zero and non-zero chemical potential. We begin by considering the case of vanishing angular momentum. For a single interval $A=\left[\left(\ell,t\right),\left(0,0\right)\right]$ located on the boundary, we deform one of the endpoints, i.e. $p_2=\left(r_c,\pm\frac{\delta}{2},\frac{\delta}{2}\right)$, while keeping the other endpoint $p_1=\left(r_c,\ell,t\right)$ fixed. 
	
	As discussed earlier, following \cref{QNEC_QFT}, we take the derivative of HEE with respect to $\delta$ and then set $\delta=0$ to obtain $S_{\text{ent}}^{\prime\prime}$. Using $T_{++}=T_{UU}$ and $T_{--}=T_{VV}$ as given in \cite{Banerjee:2024wtl}, and substituting these into \cref{QNEC_QFT}, we can determine $\mathcal{Q}_{\pm}$ for a boosted interval with $L=\bar{L}$ and $L\neq\bar{L}$. The first case corresponds to the vanishing chemical potential, while the second represents the non-vanishing chemical potential case. 
	
	Although the full non-perturbative expression is cumbersome to express analytically, for a purely timelike interval $A=\left[\left(0,t\right),\left(0,0\right)\right]$ and vanishing chemical potential, we can write the non-perturbative result as
	\begin{align}
		\mathcal{Q}_{\pm}^{\left(T\right)}=\frac{\pi ^2 c \left(\beta  \sqrt{\beta ^2-8 \pi ^2 \mu }+\left(\beta ^2-4 \pi ^2 \mu \right) \text{csch}^2\left(\frac{\pi  t}{\beta }\right)\right)}{6 \beta ^2 \left(\beta ^2-8 \pi ^2 \mu \right)}\, .
	\end{align}
	
	For the boosted interval, the perturbative expressions for QNEC become
	\begin{align}
		\mathcal{Q}_{+}&=\frac{ \pi ^2 c \csch ^2\left(\frac{\pi  (l+t)}{\beta }\right) \coth \left(\frac{\pi  (l+t)}{\beta }\right) \left(\beta  \left(\beta ^2+4 \pi ^2 \mu \right) \sinh \left(\frac{2 \pi  (l+t)}{\beta }\right)-16 \pi ^3 l \mu \right)}{12\beta ^5 }+\mathcal{O}\left(\mu^2\right)\, ,\\
		\mathcal{Q}_{-}&=\frac{ \pi ^2 c \csch ^2\left(\frac{\pi  (l-t)}{\beta }\right) \coth \left(\frac{\pi  (l-t)}{\beta }\right) \left(\beta  \left(\beta ^2+4 \pi ^2 \mu \right) \sinh \left(\frac{2 \pi  (l-t)}{\beta }\right)-16 \pi ^3 l \mu \right)}{12\beta ^5 }+\mathcal{O}\left(\mu^2\right)\, .
	\end{align}
	We observe that in this case $\mathcal{Q}_{+}\neq\mathcal{Q}_{-}$. As a consistency check, for a purely spacelike interval with $t=0$, we recover the result discussed in \cite{Banerjee:2024wtl}.
	
	The full non-perturbative QNEC expression for a timelike interval in the presence of a chemical potential is cumbersome to represent, but the perturbative expressions for the boosted interval can be written as
	\begin{align}
		\mathcal{Q}_{+}&=\frac{\pi ^2 c \coth ^2\left(\frac{\pi  (l+t)}{\beta_{+}}\right)}{6 \beta_{+}^2}
		+\frac{\pi ^4 c \mu  \coth \left(\frac{\pi  (l+t)}{\beta_{+}}\right) \text{csch}^2\left(\frac{\pi  (l+t)}{\beta_{+}}\right) \left(\beta_{-} \beta_{+} \sinh \left(\frac{2 \pi  (l+t)}{\beta_{+}}\right)-4 \pi \beta  (l -\Omega t )\right)}{3 \beta_{-}^2 \beta_{+}^4} \, ,\\
		\mathcal{Q}_{-}&=\frac{\pi ^2 c \coth ^2\left(\frac{\pi  (l-t)}{\beta_{-}}\right)}{6 \beta_{-}^2}
		+\frac{\pi ^4 c \mu  \coth \left(\frac{\pi  (l-t)}{\beta_{-}}\right) \text{csch}^2\left(\frac{\pi  (l-t)}{\beta_{-}}\right) \left(\beta_{-} \beta_{+} \sinh \left(\frac{4\beta \pi  (l-t)}{\beta_{-}}\right)-4 \pi \beta  (l -\Omega t )\right)}{3 \beta_{-}^4 \beta_{+}^2}\, .
	\end{align}
	These expressions show that, in the limit of vanishing chemical potential ($\beta_+=\beta_-$), the results match with the undeformed case.
	
	Now, we plot the real part of the undeformed HEE for a purely timelike interval and the corresponding QNEC against the deformation parameter.
	\begin{figure}[ht]
		\begin{center}
			\includegraphics[width=0.49\textwidth]{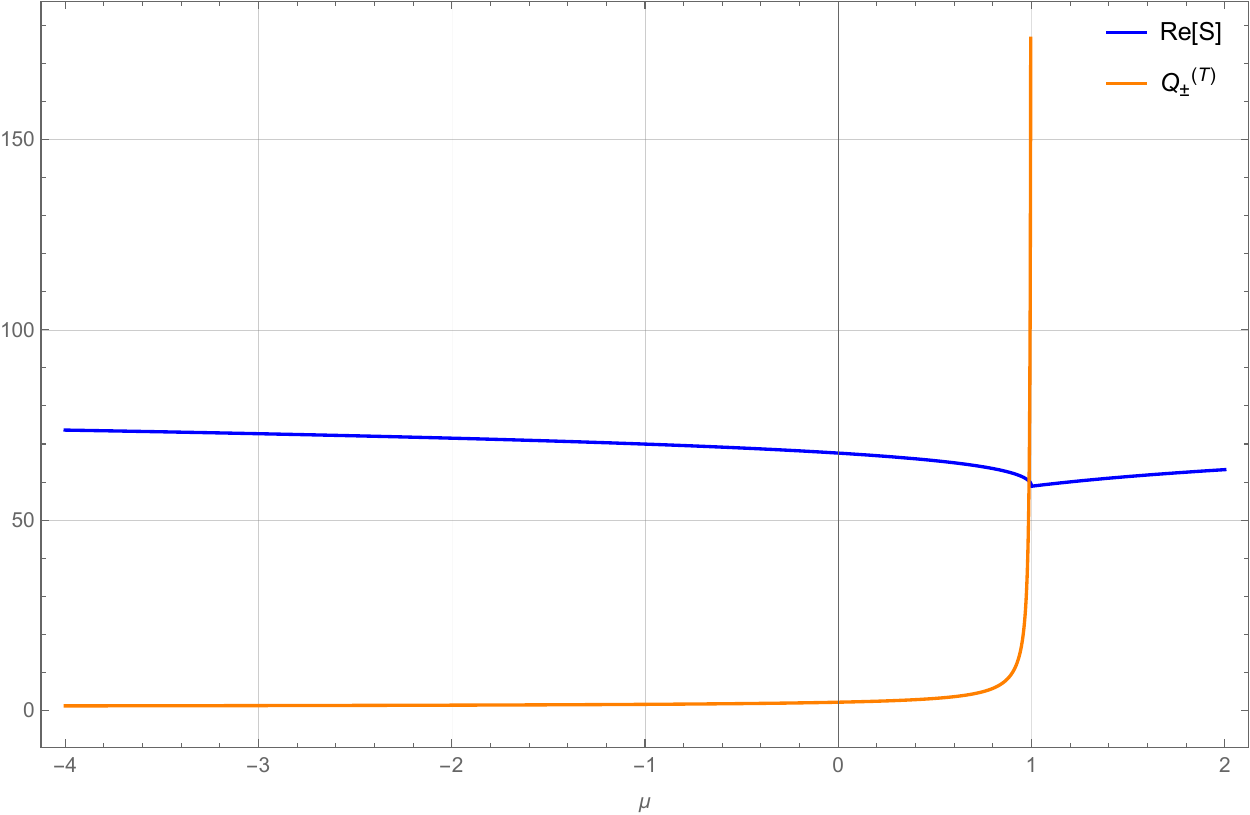}
			\includegraphics[width=0.49\textwidth]{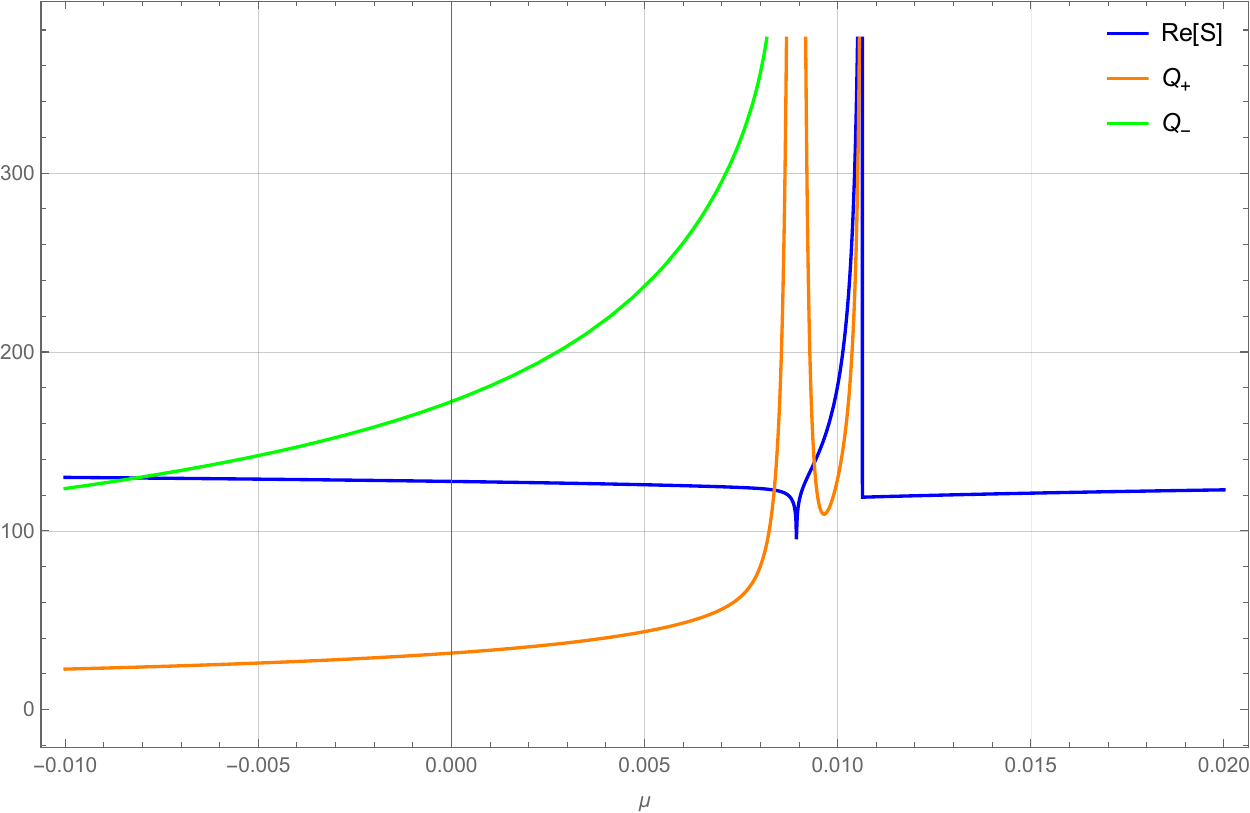}
			\caption{
				Left: Variation of the real part of HEE (blue) and $\mathcal{Q}_{\pm}$ (orange) with deformation parameter for a purely timelike interval at $t=2,~\beta=\sqrt{8}\pi,~\Omega=0,~c=12\pi,~u_c=0.01$. 
				Right: Variation of $\mathcal{Q}_{+}$ (orange) and $\mathcal{Q}_{-}$ (green) with deformation parameter for non-vanishing chemical potential at $t=2,~\beta=1,~\Omega=0.4,~c=12\pi,~u_c=0.01$.}
			\label{QvsMu}
		\end{center}
	\end{figure}
	
	In the left panel, we plot the case with vanishing chemical potential, while the right panel corresponds to non-vanishing chemical potential. The blue curve represents the real part of the HEE for a purely timelike interval. The orange curve shows $\mathcal{Q}_{\pm}$ in the left panel, while in the right panel it represents $\mathcal{Q}_{+}$ and the green curve represents $\mathcal{Q}_{-}$. We observe that in the left panel, the QNEC becomes complex beyond a certain value of $\mu$, whereas in the right panel, $\mathcal{Q}_{+}$ and $\mathcal{Q}_{-}$ exhibit distinct behaviors.
	\subsection{QNEC in the root-$\text{T}\bar{\text{T}}$ Deformation}
	In this subsection, we investigate the quantum null energy condition (QNEC) in the root-$\text{T}\bar{\text{T}}$ deformed theory. We begin with the simpler case of vanishing chemical potential (or equivalently, zero angular momentum in the bulk dual), and then move on to the finite chemical potential/rotating BTZ case. Our approach is based on computing the variation of the entanglement entropy for specific intervals and applying the general QNEC relation given in \cref{QNEC_QFT}. The $T_{++}$ and $T_{--}$ components we can collect from \cite{Ebert:2023tih}
	
	\paragraph{Zero chemical potential:}
	We first consider a boosted single interval	$A = \big[\,(\ell, t),\; (0, 0) \,\big]$ on the boundary, where one endpoint is at $p_1 = \left( r_c, \ell, t \right)$ and the other endpoint, after an appropriate shift in the null direction, is taken as $p_2 = \left( r_c, \pm \frac{\delta}{2}, \frac{\delta}{2} \right)$.
	Substituting into \cref{QNEC_QFT} yields the general (non-perturbative) expressions for $\mathcal{Q}_{\pm}$ for purely spacelike ($S$) and purely timelike ($T$) intervals:
	\begin{align}
		\mathcal{Q}_{\pm}^{(S)} &= \frac{\pi ^2 c \, e^{-\mu} \cosh (\mu) \, \coth^2\!\left( \frac{\pi  e^{-\mu} \, \ell}{\beta} \right)}{6 \beta^2} \, , \\[4pt]
		\mathcal{Q}_{\pm}^{(T)} &= \frac{\pi ^2 c \, e^{-\mu} \cosh (\mu) \, \coth^2\!\left( \frac{\pi  t}{\beta} \right)}{6 \beta^2} \, .
	\end{align}
	We note here that for purely spacelike and purely timelike intervals, the quantities $\mathcal{Q}_{+}$ and $\mathcal{Q}_{-}$ coincide exactly. This is consistent with the fact that in these symmetric configurations, the null variations in either direction are equivalent.
	
	For a boosted interval, however, the symmetry between $\mathcal{Q}_{+}$ and $\mathcal{Q}_{-}$ is broken. The non-perturbative expressions become cumbersome, so we expand perturbatively in the deformation parameter $\mu$. To leading order in $\mu$ we find
	\begin{align}
		\mathcal{Q}_{+} &= \frac{\pi^2 c \, \csch^2\!\left( \frac{\pi ( \ell + t )}{\beta} \right) 
			\coth\!\left( \frac{\pi ( \ell + t )}{\beta} \right) \left[ 4 \pi \ell \mu - \beta (\mu - 1) \sinh\!\left( \frac{2\pi (\ell + t)}{\beta} \right) \right]}{12 \beta^3} + \mathcal{O}(\mu^2) \, , \\[4pt]
		\mathcal{Q}_{-} &= \frac{\pi^2 c \, \csch^2\!\left( \frac{\pi ( \ell - t )}{\beta} \right) 
			\coth\!\left( \frac{\pi ( \ell - t )}{\beta} \right) \left[ 4 \pi \ell \mu - \beta (\mu - 1) \sinh\!\left( \frac{2\pi (\ell - t)}{\beta} \right) \right]}{12 \beta^3} + \mathcal{O}(\mu^2) \, .
	\end{align}
	From these results we clearly see that $\mathcal{Q}_{+} \neq \mathcal{Q}_{-}$ for generic boosted intervals, as the arguments $(\ell + t)$ and $(\ell - t)$ appear differently. Physically, this difference arises because the boost mixes space and time directions, breaking the symmetry between left-moving and right-moving null variations.
	
	Here we plot the variation of QNEC with the deformation parameter, and compare with the HEE on the same plot for different types of subsystems.
		\begin{figure}[ht]
		\begin{center}
			\includegraphics[width=0.49\textwidth]{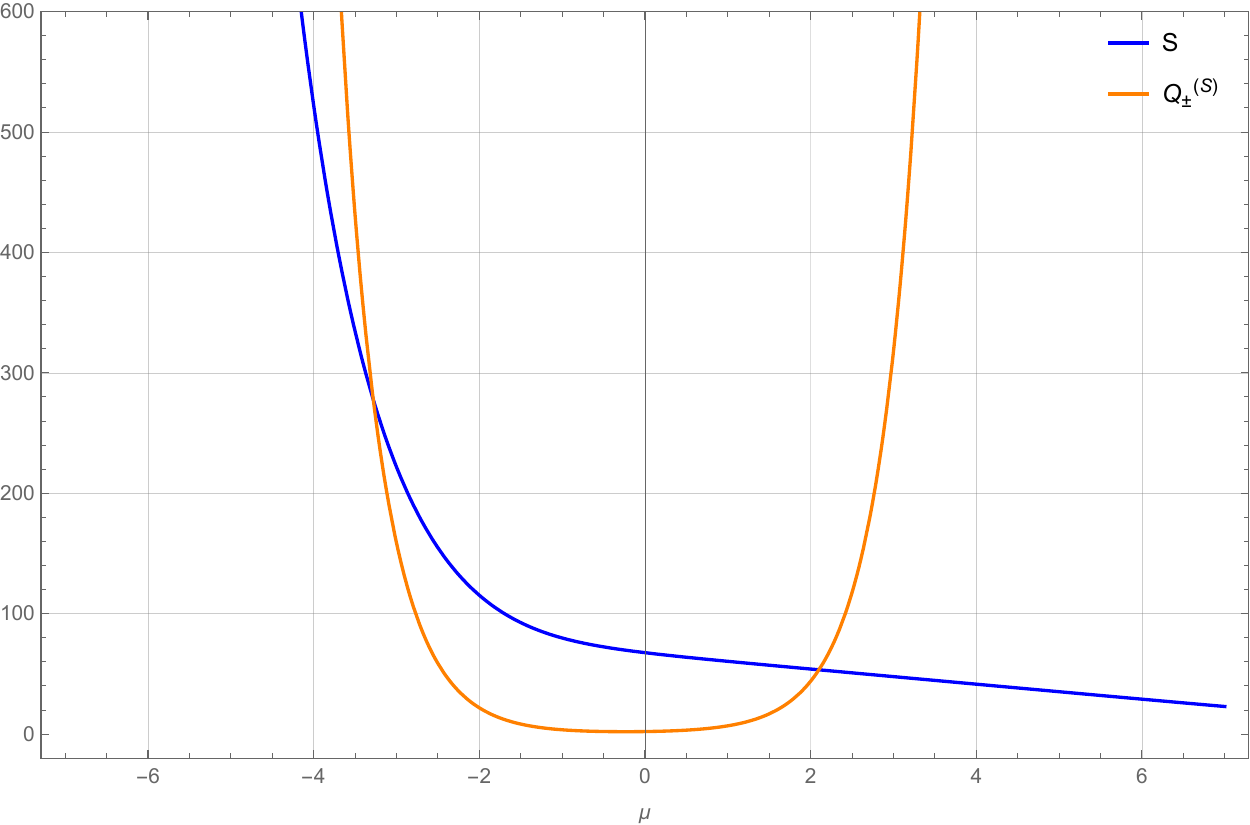}
			\includegraphics[width=0.49\textwidth]{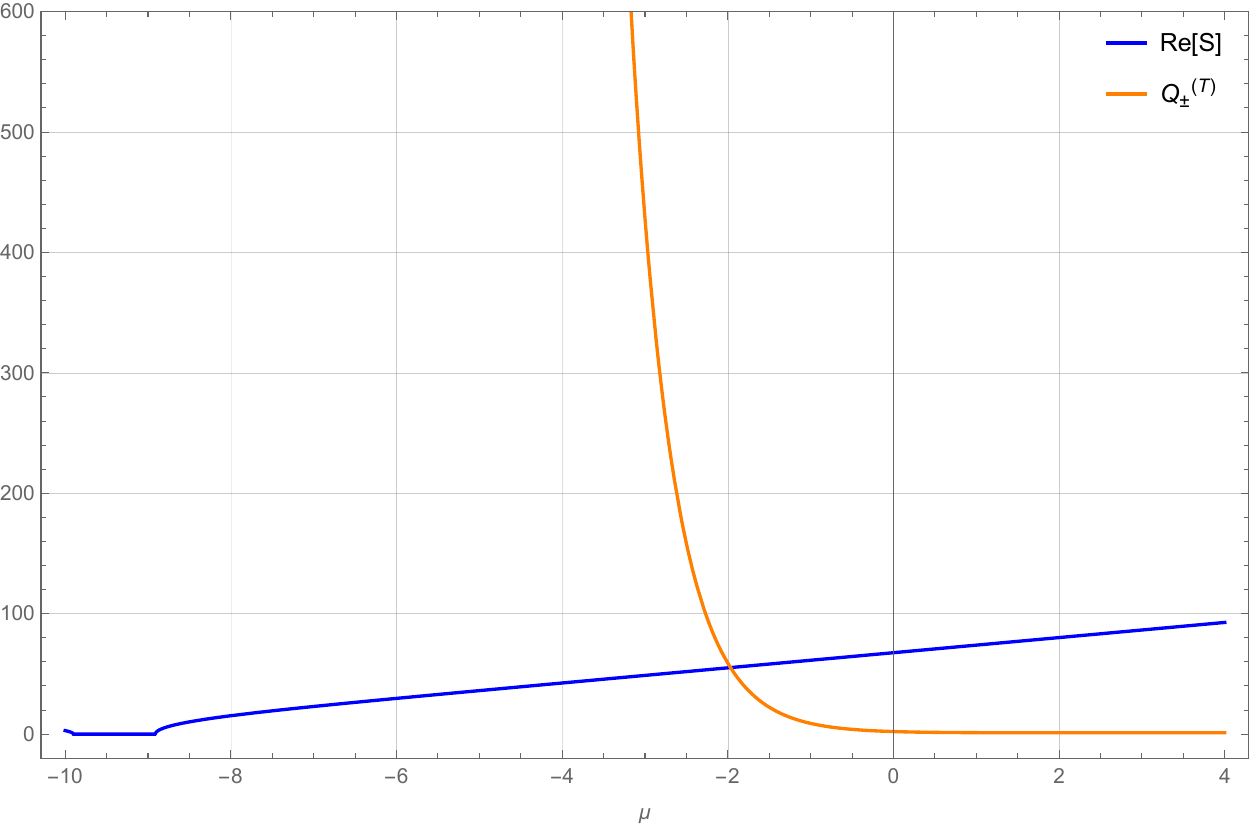}
			\caption{Left panel $t=2,\beta=\sqrt{8}\pi,\Omega=0,c=12\pi, u_c=0.01$, Right panel  $t=2,\beta=\sqrt{8}\pi,\Omega=0,c=12\pi, u_c=0.01$}
			\label{QvsMu}
		\end{center}
	\end{figure}
	On the left panel, we plot the HEE for a purely spacelike interval, where the blue curve represents the HEE and the orange curve represents $\mathcal{Q}_{\pm}$. We can observe almost a symmetric behaviour for the QNEC; it is real and positive for any deformation parameter. On the right panel, we can see the behaviour of QNEC for a purely timelike interval, where the blue curve represents the real part of timelike HEE and the orange curve represents $\mathcal{Q}_{\pm}^{(T)}$.
	
	\paragraph{Finite chemical potential :}
	Next, we consider a boosted interval $A$ in the rotating BTZ background, which corresponds to a finite chemical potential in the dual field theory. The setup is similar to the previous case: we shift one endpoint of the interval along the null direction as before, compute the entanglement entropy using the appropriate rotating BTZ geodesic length formula, and apply \cref{QNEC_QFT}. Here, the $T_{\pm\pm}$ components are not equal, and they are considered from \cite{Ebert:2023tih}.
	
	As in the boosted case above, the exact non-perturbative expression is lengthy, so we expand to first order in $\mu$. For the $+$ and $-$ null directions we find
	\begin{align}
		\mathcal{Q}_{+} &= \frac{\pi^2 c \, \coth^2\!\left( \frac{\pi (\ell + t)}{\beta_+} \right)}{6 \beta_+^2} \;+\;
		\frac{\pi^2 c \, \mu \, \coth\!\left( \frac{\pi (\ell + t)}{\beta_+} \right) \, \text{csch}^2\!\left( \frac{\pi (\ell + t)}{\beta_+} \right) 
			\left[ 4 \pi \beta (\ell - \Omega t) - \beta_- \beta_+ \sinh\!\left( \frac{2\pi (\ell + t)}{\beta_+} \right) \right]}{12 \beta_- \beta_+^3} \, , \\[4pt]
		\mathcal{Q}_{-} &= \frac{\pi^2 c \, \coth^2\!\left( \frac{\pi (\ell - t)}{\beta_-} \right)}{6 \beta_-^2} \;+\;
		\frac{\pi^2 c \, \mu \, \coth\!\left( \frac{\pi (\ell - t)}{\beta_-} \right) \, \text{csch}^2\!\left( \frac{\pi (\ell - t)}{\beta_-} \right) 
			\left[ 4 \pi \beta (\ell - \Omega t) - \beta_- \beta_+ \sinh\!\left( \frac{2\pi (\ell - t)}{\beta_-} \right) \right]}{12 \beta_-^3 \beta_+} \, .
	\end{align}
	Here $\beta_+$ and $\beta_-$ are the inverse temperatures associated with the left- and right-moving sectors of the rotating BTZ black hole, and $\Omega$ is the angular velocity of the horizon. The difference between $\mathcal{Q}_{+}$ and $\mathcal{Q}_{-}$ is now controlled both by the boost and by the rotation, and encodes the interplay between null deformations and the chiral asymmetry of the thermal state.
	
	Though we can not express the full non-perturbative expression for the $\mathcal{Q}_{\pm}$, we can do numerical analysis and plot it with the deformation parameter. The left panel represents a spacelike interval and the right panel represent for a timelike interval. 
			\begin{figure}[ht]
		\begin{center}
			\includegraphics[width=0.49\textwidth]{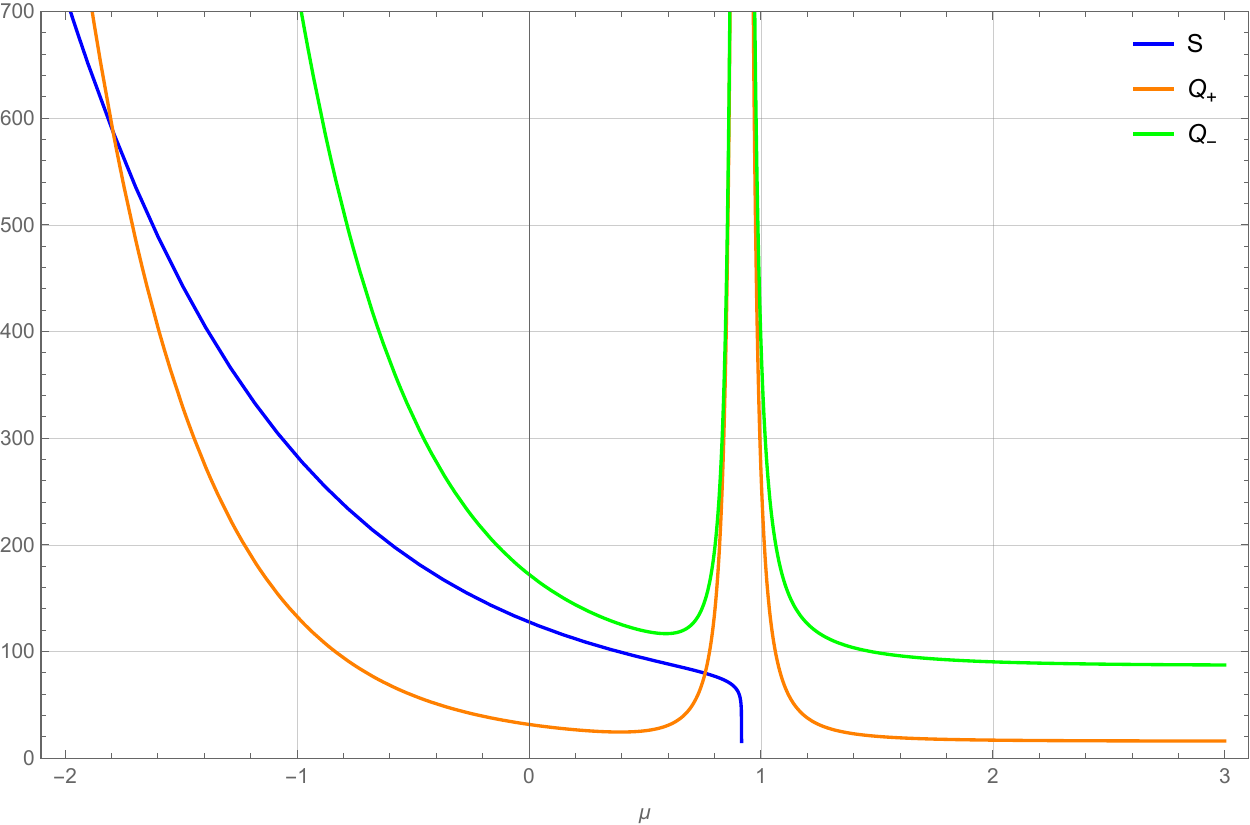}
			\includegraphics[width=0.49\textwidth]{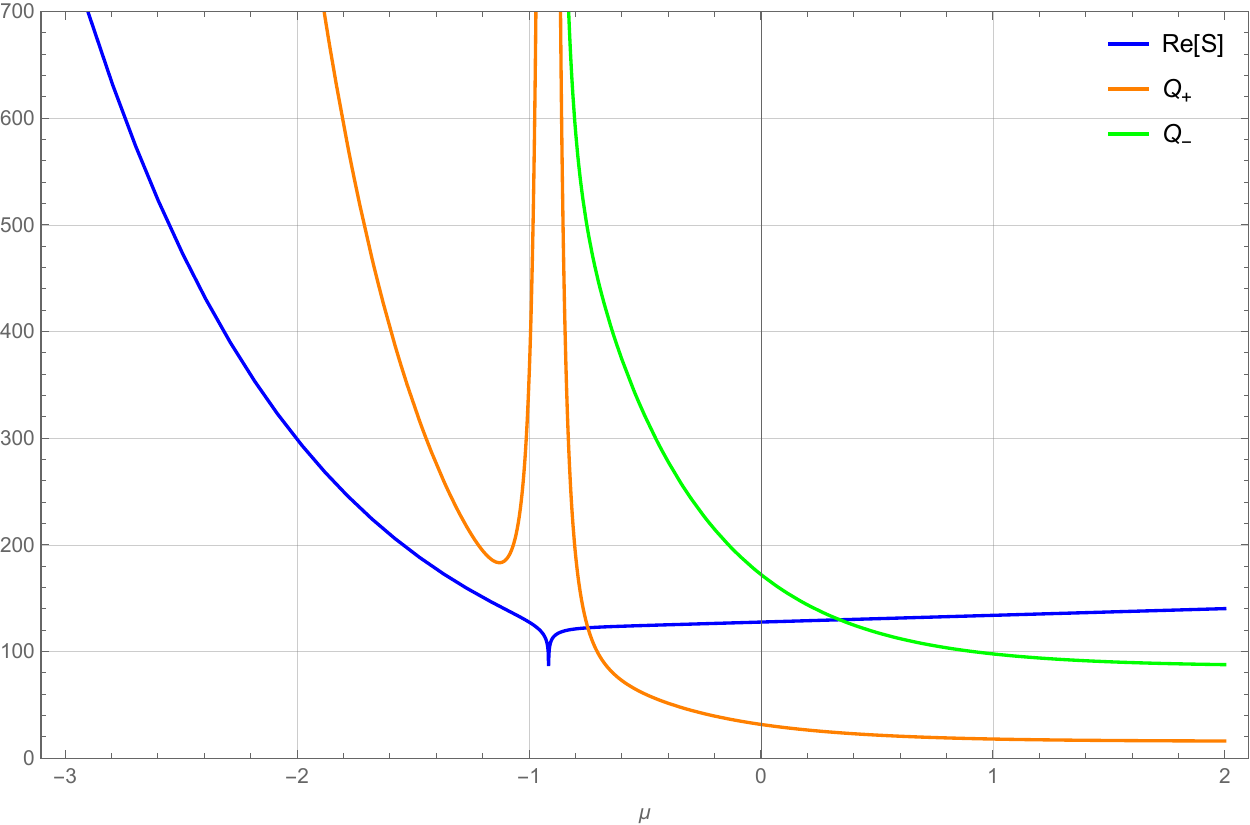}
			\caption{Left panel $l=2,\beta=1,\Omega=0.4,c=12\pi, u_c=0.01$, Right panel  $t=2,\beta=1,\Omega=0.4,c=12\pi, u_c=0.01$}
			\label{QvsMu}
		\end{center}
	\end{figure}
	
	On the left panel, the blue line is for the HEE, and the orange and green lines represent $\mathcal{Q}_+$ and $\mathcal{Q}_-$ respectively. On the right panel, the blue line represents the real part of timelike HEE, and the orange and green line represents $\mathcal{Q}_+$ and $\mathcal{Q}_-$.
	 \section{Summary}\label{sec7}
  In this work, we studied the impact of $\text{T}\bar{\text{T}}$ and root-$\text{T}\bar{\text{T}}$ deformations on information-theoretic measures within the holographic AdS$_3$/CFT$_2$ framework, focusing on BTZ and rotating BTZ black hole geometries in the bulk. The CFT dual to the BTZ background corresponds to a thermal CFT, while the rotating BTZ black hole is dual to a thermal CFT with a conserved charge. Using the mixed boundary condition approach, we computed the leading corrections to a mixed-state entanglement measure, namely the reflected entropy, for bipartite states. 

    Furthermore, we analyzed the modification of entanglement structure induced by the root-$\text{T}\bar{\text{T}}$ deformation for a single-interval bipartite pure state, considering both vanishing and non-vanishing conserved charge in CFT$_2$, with the subsystem subjected to a Lorentz boost. In the presence of a conserved charge, the CFT is defined on a twisted cylinder, corresponding to a compactification along an arbitrary direction. For computational convenience, we employed the coordinate rotation technique available in the literature. On the gravity side, we also examined the corresponding deformed bulk geometries and analyzed how the deformation modifies the relevant geodesic lengths.

 The CFT duals of the three-dimensional BTZ and rotating BTZ black hole geometries are, respectively, a thermal CFT$_2$ and a thermal CFT$_2$ with a conserved charge. For these backgrounds, our results for the deformed entanglement wedge cross section agree, up to first order in the deformation parameter $\mu$, with one half of the reflected entropy corrections obtained in Refs.~\cite{Basu:2024bal,Basu:2024enr}. Remarkably, additional first-order corrections to the EWCS for different bipartite subsystems in these geometries have been computed using the cut-off prescription, and our results are consistent with those findings as well, to the same perturbative order \cite{Basu:2024bal,Basu:2024enr}. 

    Moreover, we investigated the entanglement entropy for purely timelike intervals. We uncovered an intriguing feature: for a specific range of negative values of the deformation parameter $\mu$, the timelike entanglement entropy remains real. Such behavior does not arise in the cut-off formulation, which is restricted to positive values of $\mu$. In the second part of this work, we computed the correction to the entanglement entropy of a Lorentz-boosted subsystem induced by the root-$\text{T}\bar{\text{T}}$ deformation using conformal perturbation theory. 

    On the gravity side, we first constructed the corresponding deformed bulk geometry and then applied the Ryu--Takayanagi prescription to compute the holographic entanglement entropy. We explicitly evaluated the associated geodesic lengths and found agreement with the field-theoretic results up to first order in the deformation parameter. Analyzing the non-perturbative bulk expressions further revealed that, in the presence of non-vanishing angular momentum, the theory imposes an upper bound on the deformation parameter for spatial intervals. For purely timelike intervals, a behavior analogous to that observed in the mixed boundary condition analysis of the $\text{T}\bar{\text{T}}$ deformation emerges: within a specific negative range of the deformation parameter, the non-perturbative entanglement entropy remains positive. Finally, using the non-perturbative holographic entanglement entropy, we analyzed the validity of the quantum null energy condition.

 In this work, we investigated BTZ and rotating BTZ, which have a physical temperature, whereas an extremal BTZ black hole does not have any physical temperature. In such a scenario, the $\text{T}\bar{\text{T}}$ and root-$\text{T}\bar{\text{T}}$ deformed bulk geometric structure might be an interesting direction to explore. If we examine the deformed rotating BTZ geometry, we can observe setting both the radii equal blows up the metric components, as it is interesting to observe the fact behind this behavior. Although we can take the extremal limit from the non-extremal result, it will be much more convenient to study both from the field theory and bulk perspectives. In \cite{Banerjee:2024wtl}, it was argued that the upper bound on the deformation parameter is set by the Hagedorn bound, beyond which thermal equilibrium ceases to exist. For timelike intervals, this upper bound remains unchanged when the entanglement entropy is required to be real. In the case of the root-$\text{T}\bar{\text{T}}$ deformation, the corresponding bound differs and depends solely on the angular momentum of the rotating BTZ black hole, for both spacelike and timelike intervals. Adopting a similar perspective for reflected entropy, one may anticipate analogous bounds on the deformation parameter $\mu$ for both deformations. However, ensuring the reality of the EWCS in the deformed geometry makes it considerably more subtle to comment on the Hagedorn bound from the viewpoint of mixed-state measures.

 The interesting future directions related to our study might be to investigate the other entanglement mixed state measures like entanglement negativity \cite{Vidal:2002zz,Plenio:2005cwa}, the entanglement of purification \cite{Takayanagi:2017knl}, and the balanced partial entanglement \cite{Wen:2021qgx}. The mixed state entanglement corrections in field theory are also an important direction to explore. In our root-$\text{T}\bar{\text{T}}$ field theoretic analysis, our approach faces difficulties in handling more than one interval. It would be exciting to develop a similar $\text{T}\bar{\text{T}}$ like technique  \cite{Jeong:2019ylz,Kudler-Flam:2020url,Basu:2023aqz,Basu:2023bov,Basu:2024bal,Basu:2024enr} for several other mixed state measures, and finding the deformed bulk structure from root-$\text{T}\bar{\text{T}}$ deformed AdS geometries. Some other very interesting issues related to $\text{T}\bar{\text{T}}$ and root-$\text{T}\bar{\text{T}}$ deformation are to be found beyond the leading order corrections from the field theoretic perspective. One can also explore the effects of these deformations on other kinds of CFT, like BCFT and its dual AdS.
 
 Finally, we conclude that our work is consistent with the AdS/CFT correspondence for both $\text{T}\bar{\text{T}}$ and root-$\text{T}\bar{\text{T}}$ deformed thermal CFTs, and thermal CFTs with conserved charge. Our investigation offers several insights that may be useful for future studies of entanglement and deformed theories.
	
	\section*{Acknowledgment}
	 We would like to thank Debarshi Basu, Ankit Anand, and Ankur Dey for helpful discussions and insightful suggestions that contributed to the progress of this work. SB also acknowledges the financial support from the Council of Scientific and Industrial Research (CSIR) under grant number 09/0092(12686)/2021-EMR-I.
	\appendix
	\section*{Timelike EE in $\text{T}\bar{\text{T}}$ deformed MBC framework with non zero chemical potential }
	Here we consider a purely timelike interval $A=\left[\left(0,t\right),\left(0,0\right)\right]$. To find the holographic entanglement entropy in the bulk rotating BTZ geometry deformed by $\text{T}\bar{\text{T}}$ operator, we simply use the Ryu-Takayanagi formula and the embedding coordinates given in \cref{Embedding-coordinates}, and finally, we are left with the HEE as follows
	\begin{align}
		S^{\left(T\right)}=\frac{1}{4G}\log \left(
\frac{
\left(8 \pi ^2 \mu ^2\right)
\left(
\cosh \left(
\frac{\pi t (\beta_+-\beta_-)}
{\beta_- \beta_+ \sqrt{1-\frac{8 \pi ^2 \mu }{\beta_- \beta_+}}}
\right)
-
\cosh \left(
\frac{\pi t (\beta_-+\beta_+)}
{\beta_- \beta_+}
\right)
\right)
}{
\mathrm{u_c}^2 \left(
\beta_- \beta_+
\left(
1-\sqrt{1-\frac{8 \pi ^2 \mu }{\beta_- \beta_+}}
\right)^2
\right)
}
\right)\, ,\label{TTbar_EE2}
	\end{align}
	here, $u_c$ is related with the IR cut-off $u_c\sim\frac{1}{r_c}$. One thing to note in this result it might be real depending on the deformation parameter $\mu$. If we investigate the, the square root term is real for 
	\begin{align}
		\mu\le\frac{\beta_+\beta_-}{8\pi^2}\, .
	\end{align}
	Further, if we imply the argument of the logarithm has to be positive, then one can easily observe that the differences of the two $cosh$ terms have to be positive. From there, we can conclude if we have the condition 
	\begin{align}
		\mu \ge\frac{\beta^2 \left(1-\Omega^2\right)^2}{8\pi^2}\, ,
	\end{align}
    then the argument inside the $log$ will be positive. Therefore we finally get a window for which it is observed that the total non-perturbative entropy can be real for those values of $\mu$ which satisfies the condition as follows
	\begin{align}
		\frac{\beta_+\beta_-}{8\pi^2}\ge\mu \ge\frac{\beta^2 \left(1-\Omega^2\right)^2}{8\pi^2}\, .
	\end{align}
    From here we can see, for a non extremal black hole the window stays away from zero, therefor for small deformation parameter, the geodesic length can not be real. Another important thing to note here, as $\Omega\to0$ the window closes. We can say that for a non-rotating BTZ, we do not see such behavior.
	
	   Using this reault, and further doing a perturbation expansion w.r.t $\mu$, it can be verified that the first order correction matches with the CFT result obtained in \cite{Basu:2024enr}.
	\bibliographystyle{utphys}
	\bibliography{reference}
\end{document}